\documentclass[twocolumn]{jpsj2} 
\title{Interplay between carrier localization and magnetism in diluted magnetic and ferromagnetic semiconductors}

\author{Tomasz \textsc{Dietl}$^{1,2,3}$\thanks{E-mail address: dietl at ifpan.edu.pl}}

\inst{$^{1}$Institute of Physics, Polish Academy of Science, al.~Lotnik\'ow 32/46, PL 02-668 Warszawa, Poland\\
$^{2}$Institute of Theoretical Physics, University of Warsaw, PL-00-681 Warszawa, Poland\\
$^{3}$ ERATO Semiconductor Spintronics Project, Japan Science and Technology Agency, al.~Lotnik\'ow 32/46, PL-02-668 Warszawa, Poland}

\abst{The presence of localized spins exerts a strong influence on quantum localization in doped semiconductors. At the same time  carrier-mediated interactions between the localized spins are modified or even halted by carriers' localization. The interplay of these effects is discussed for II-VI and III-V diluted magnetic semiconductors. This insight is exploited to interpret the complex dependence of resistance on temperature, magnetic field, and concentration of valence-band holes in (Ga,Mn)As. In particular, high field negative magnetoresistance results from the orbital weak localization effect. The resistance maximum and the associated negative magnetoresistance near the Curie temperature are assigned to the destructive influence of preformed ferromagnetic bubbles on the "antilocalization" effect driven by disorder-modified carrier-carrier interactions. These interactions account also for the low-temperature increase of resistance. Furthermore, the sensitivity of conductance to spin splitting and to scattering by spin disorder may explain resistance anomalies at coercive fields, where relative directions of external and molecular fields change.}

\kword{diluted magnetic semiconductors, carrier-mediated ferromagnetism, quantum localization}

\begin{document}
\maketitle

\section{Introduction}

In course of the years (Ga,Mn)As has reached the status of a model system for understanding\cite{Matsukura:2002_c,Jungwirth:2006_a} and expoliting\cite{Dietl:2007_c} carrier-controlled ferromagnetism in semiconductors. One of the prominent aspect of this system is the interplay between carrier-mediated ferromagnetism and carrier localization. In this paper we first recall the present view on the nature of localization in doped semiconductors. We then discuss results of  comprehensive studies of interplay between localization and magnetism in doped II-VI diluted magnetic semiconductors. Equipped with relevant information, we present mechanisms which determine charge transport properties of ferromagnetic (Ga,Mn)As films.

\section{Localization in non-magnetic doped semiconductors}
\subsection{Critical concentration}

The metal-insulator transition (MIT) is one of most characteristic features of doped semiconductors.\cite{Belitz:1994_a} In the bulk three-dimensional (3D) case the zero-temperature conductivity vanishes at $r_s \approx 2.5$, where $r_s$ is the ratio of an average distance between the carriers to the Bohr radius $a_B$ of the relevant dopant, $r_s = (3/4\pi n)^{1/3}/a_B$, with $n$ being the carrier concentration or, equivalently, the net concentration of the majority impurities. Accordingly, the critical carrier concentration $n_c$, where $n_c^{1/3}a_B \approx 0.25$, is of the order of $10^{14}$~cm$^{-3}$~in \textit{n}-InSb but rises to $10^{16}$ and $10^{19}$~cm$^{-3}$ for GaAs doped with shallow donors and acceptors, respectively. Obviously, $a_B$  and thus $n_c$ can be changed by, {\em e.~g.} strain or the magnetic field.

\subsection{Beyond the region of metal-insulator transition}

At low carrier densities, $n \ll n_c$, charge transport proceeds due to thermally activated phonon-assisted hopping between impurity states. In this regime many-body correlation effects are of primary importance making, for instance, that the particular impurities are occupied only by a single carrier and that the one-electron density-of-states acquires a gap at the Fermi energy.\cite{Shklovskii:1984_a}

In the high density range, $n \gg n_c$, many-body screening washes out bound states, so that carriers abandon parent impurities and reside in the relevant band.  Here, carrier correlations and disorder-induced band tailing shift the band down in energy by $\Delta E_g \approx 2E_I/r_s$, where $E_I$ is the impurity binding energy. This sizable band-gap narrowing is clearly seen in tunneling and luminescence experiments.\cite{Palankovski:1999_a} At the same time, according to Landau theory of Fermi liquids, many-body effects result in a minor renormalization of the carrier dispersion for $n \gg n_c$. Furthermore, since screened Coulomb potentials are weak, the conductivity is well described by the Drude-Boltzmann model in this regime. A number of experiments, including Shubnikov de Haas oscillations, magnetoplasma resonances, and thermoelectric effects support the applicability of this simple approach to charge and heat transport in the region of high densities.

\subsection{Quantum localization}

The experimentally relevant impurity concentrations correspond rather often to the transition region between the two extreme situations described above. Accordingly, neither single impurity wave functions nor Bloch-type plane waves can serve for the carrier description. In the metallic regime, single-particle and many-body quantum interference effects leading to Anderson-type and Mott-type localization, respectively, are of crucial importance. These quantum effects modify the Drude-Boltzmann conductance and ultimately lead to MIT on decreasing the carrier density towards $n_c$.

However, MIT may also be viewed as delocalization of carriers residing in the impurity band. Indeed,  with the increase of the impurity concentration, an overlap between the single-impurity wave functions increases, which can ultimately lead to an insulator-to-metal transition. Within this picture, the MIT occurs before the impurity states merge with the relevant band. Accordingly, there is a region of concentrations, for which metallic-like charge transport takes place within the impurity band, whose characteristics have little to do with those of unperturbed band states. Arguments summarized below show that this is {\em not} the case. Actually, MIT in doped semiconductors appears to be of the Anderson-Mott character, {\em i.~e.}, occurs primarily due to localization of band carriers by scattering. This means, as argued in details recently,\cite{Jungwirth:2007_a,Ohno:2007_a} that states relevant for the carrier-mediated ferromagnetism retain the band character across MIT.

\subsection{Two fluids model}

It is  a formidable task to describe quantitatively effects of both disorder and carrier-carrier correlations near the Anderson-Mott transition.\cite{Belitz:1994_a} In particular, standard computational tools, such as the coherent potential approximation (CPA), do not capture the relevant physics. A  large number of studies in the past have led to the conclusion that doped semiconductors  near the MIT exhibit duality, {\em i.~e.}, they show metallic band-like nature in one type of measurements, whereas at the same time they can exhibit impurity-band-like nature in another.  There were observations of 1$s$-to-2$p$ impurity transitions in metallic $n$-GaAs,\cite{Liu:1993_a} valley splitting effects in $n$-Ge,\cite{Rosenbaum:1990_a} and the presence of a Curie-Weiss component in the magnetic susceptibility.\cite{Paalanen:1988_a} All of these can be interpreted as evidence of transport taking place in the impurity band. On the other hand, weak localization theory\cite{Altshuler:1985_a,Fukuyama:1985_a,Lee:1985_a} which assumes that localization has Anderson's character, {\em i.~e.} localization of band carriers by scattering, is successful in quantitatively explaining the temperature and field dependence of the conductance. Specific heat and Pauli susceptibility measurements, for example in $n$-Si, can be described by assuming a band mass across MIT.\cite{Paalanen:1988_a} In this situation, the two-fluid model of electronic states in doped semiconductors has been proposed.\cite{Paalanen:1991_a}

\subsection{Weak localization regime and metal-to-insulator transition}

In a somewhat simplified picture, the co-existence of two kinds of behavior discussed above results from the fact that randomness allows for the presence of isolated impurities, whose strong Curie-like paramagnetism and optical response dominate in some temperature and spectral regions, respectively, even if their concentration is statistically irrelevant. Within this view, the statistically relevant states are band-like across the MIT. As mentioned above, the Anderson-Mott localization of such states can occur owing to two quantum phenomena specific to many-body disordered systems.\cite{Belitz:1994_a,Altshuler:1985_a,Fukuyama:1985_a,Lee:1985_a} First is  one-electron interference of scattered waves corresponding to self-intersecting trajectories. Second results from interference of carrier-carrier interaction amplitudes corresponding to successive carrier-carrier scattering events that happen due to the diffusive character of carriers' motion in disorder systems. Importantly, if spin effects are taken into account, each of these two quantum effects produces corrections of both signs ("localization" and "anti-localization" terms, respectively), whose relative importance depends on the presence of perturbations affecting spin and phase of the wave functions. Hence, the conductivity near MIT shows a strong and specific dependence on the dimensionality, magnetic field, spin splitting, spin-dependent scattering, temperature, and frequency. In particular, a fitting of the a.~c. conductivity values by the Drude formula may lead to highly misleading results. These quantum effects control also a character of the critical behavior at MIT as well as account for universal conductance fluctuations and quantum noise in diffusive nanostructures.

In parallel to $r_s$, it is convenient to describe the degree of localization by the the product of the Fermi wave vector $k_F$ and mean free path $l$ calculated for given scattering potentials within the standard lowest order perturbation theory. Actually, the parameter $k_Fl$ is more universal than $r_s$, and characterizes the Anderson-Mott localization for an arbitrary form of static disorder. According to the Brooks-Herring formula for the case of ionized impurity scattering $k_Fl$ and $r_s$ are related in a simple relation, which at MIT, {\em i.~e.}  $r_s = 2.5$, gives $k_Fl \approx 5.8(1-K)/(1+K)$, where $0<K<1$ is the compensation ratio.  In particular, the Drude-Boltzmann theory is valid in the limit $k_Fl \gg 1$.

When disorder increases, so that the magnitude of $k_Fl$ decreases, we enter into the so-called weakly localized regime. Here quantum corrections start to affect the conductivity value significantly so that $\sigma$ ceases to be related to the microscopic parameter $l$ in the standard way, $\sigma = e^2k_F^2l/(3\pi^2\hbar)$.

\begin{figure}[tb]
\begin{center}
\includegraphics{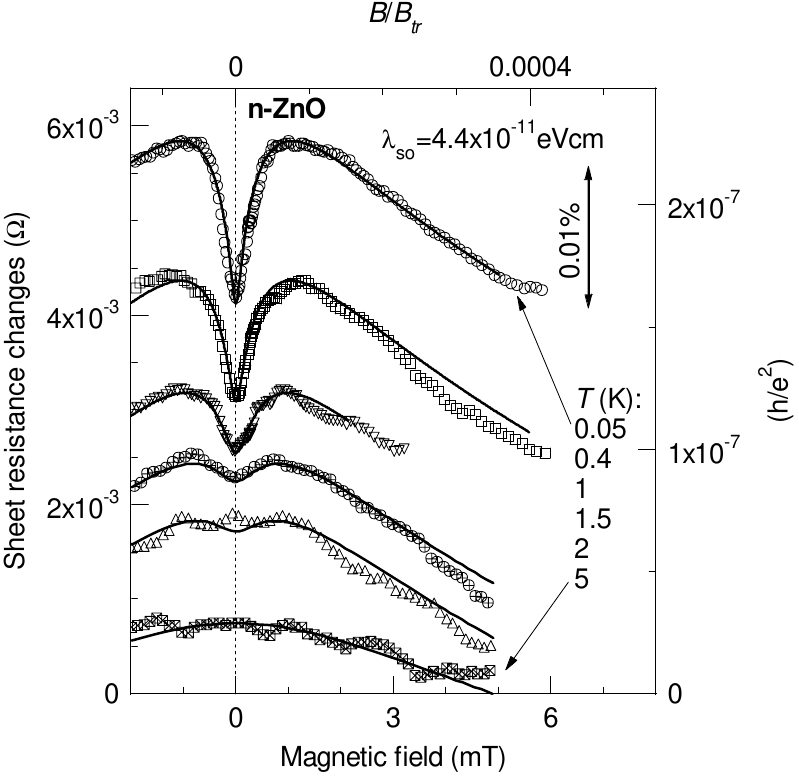}
\end{center}
\caption{Sheet resistivity changes in the magnetic field \emph{B} for thin film of \textit{n}-ZnO (symbols) compared to calculations (solid lines) within the weak localization theory for 2D. Curves are vertically shifted for clarity. At upper axis the magnetic field is
normalized to $B_{tr}\equiv\hbar/4eD\tau=14$ T (after Andrearczyk {\em et al.}\cite{Andrearczyk:2005_a} reproduce with the permission, Copyright (2005) by the American Physical Society).}
\label{f1}
\end{figure}

The presence of quantum corrections to conductivity has been verified quantitatively in a number of doped semiconductors. One of many examples is shown in Fig.~1, where magnetoresistance of a \textit{n}-ZnO thin film  is shown for various temperatures.\cite{Andrearczyk:2005_a} As seen, in the weakly localized regime ($k_Fl \approx 10$ for this sample), the present theory\cite{Altshuler:1985_a,Fukuyama:1985_a,Lee:1985_a} describes quite precisely a rather complex field and temperature dependence of resistivity. The quantitative description of the data allowed one to determine the magnitude of the Rashba spin-orbit term specific to the wurtzite ZnO, $\lambda_{so}$, and the phase breaking time brought about by inelastic scattering processes. It worth recalling that the weak-field positive magnetoresistance (MR) results from the one-particle spin-orbit "anti-localization" effect, while negative MR at higher fields has the orbital origin -- it comes from the influence of the magnetic field $B$ (vector potential) on the phase difference of two transmission amplitudes corresponding to two possible pathes along self-crossing trajectories. The corresponding field-induce conductance increase, $\Delta \sigma(B) =  0.605[e^2/(2\pi^2\hbar)](eB/\hbar)^{1/2}$ in the 3D case and two subbands, is a remarkably universal phenomenon, showing up independently of the degree of carrier-liquid spin-polarization, provided that the cyclotron frequency $\omega_c$ is greater than $1/(k_Fl\tau_i)$, where $1/\tau_i$ is a sum  of scattering rates corresponding to spin-orbit, spin-disorder, and phase breaking inelastic scattering processes.

\begin{figure}[tb]
\begin{center}
\includegraphics[scale=1.3]{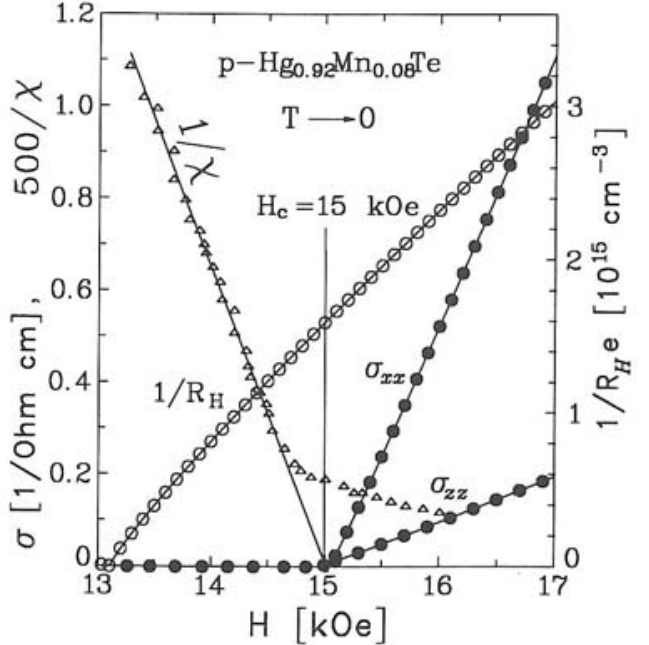}
\end{center}
\caption{Critical behavior at the insulator-to-metal transition induced by the magnetic field  (spin splitting) in paramagnetic \textit{p}-(Hg,Mn)Te. The data were obtained extrapolating results of measurements between 30 and 800~mK. The magnetic field was applied along $z$ direction. Critical exponents of static dielectric constant $\chi$ and of the conductivity tensor components $\sigma_{xx}$ and $\sigma_{zz}$ are consistent with the expectation for the transition in a spin polarized band. No critical behavior of the Hall coefficient $R_H$ is expected, and the vanishing of $1/R_H$ is assigned to the disappearance of the Fermi liquid-like states  deeply in the insulator phase (after Jaroszy\'nski {\em et al.}\cite{Jaroszynski:1992_a} reproduce with the permission, Copyright (1992) by Elsevier).}
\label{f2}
\end{figure}

When the value of $k_Fl$ decreases, the critical region is reached, occurring typically at  $k_Fl < 2$. According to the scaling theory of Anderson-Mott localization, the quantum interference effects make the value of zero-temperature conductance to vanish in a continues way on approaching  MIT. An example of such a behavior is presented in Fig.~2 for p-(Hg,Mn)Te,\cite{Wojtowicz:1986_a,Jaroszynski:1992_a} in which the insulator-to-metal transition is driven by the field-induced ordering of the Mn spins.  On the insulator side of MIT, it is the inverse of the carrier localization length which is gradually and continually increasing from zero at MIT to the inverse of the Bohr radius deeply in the strongly localized regime. The divergence of the localization length on approaching the critical point from the insulator side of MIT is witnessed by the critical behavior of static dielectric function and hopping conductivity, as shown in Figs.~2 and 3.\cite{Wojtowicz:1989_a,Jaroszynski:1992_a}

\begin{figure}[tb]
\begin{center}
\includegraphics[scale=1.8]{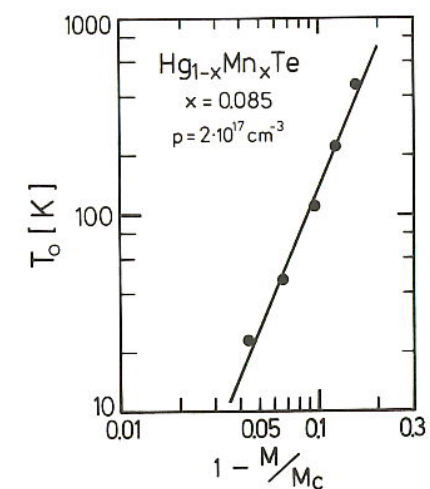}
\end{center}
\caption{Scaling of activation energy $T_o$ of variable range hopping conductivity in paramagnetic \textit{p}-(Hg,Mn)Te on the insulator side of the field-induced metal-to-insulator transition (cf. Fig.~2). The transition is driven by magnetization-induced spin splitting, and occurs at magnetization $M_c$. The experimental value of the critical exponent 2.4 (slope of the solid line) agrees with the expectations for the spin-polarized universality class (after Wojtowicz {\em et al.}\cite{Wojtowicz:1989_a} reproduce with the permission, Copyright (1989) by Elsevier).}
\label{f3}
\end{figure}

\section{Effects of magnetic ions on localization}
\subsection{Trapping and doping}

A number of transition metal and rare earth impurities gives rise to deep band-gap levels derived from highly localized $d$ and $f$ shells. Owing to the relatively large on-site correlation energy $U$, the acceptor state (which can trap an electron) resides at much higher energies than the donor state (which can donate an electron or in other words trap a hole).  Magnetic impurities that introduce mid-gap states are exploited to fabricate semi-insulating materials, such as GaAs:Cr or InP:Fe. In contrast, if the impurity levels of the donor or acceptor character are degenerate with the conduction or valence band, respectively, shallow donor-like or acceptor-like states appear in the band gap. To this class belong Sc in CdSe\cite{Glod:1994_a} and Mn in GaAs,\cite{Matsukura:1998_a} which  exhibit properties specific to \textit{n}-type and \textit{p}-type semiconductors, respectively. Furthermore, it has been suggested that a strong local potential associated with the magnetic atom can enhance the biding energy (by the so-called central cell corrections) or even produce a bound state in the gap.\cite{Benoit:1992_a,Dietl:2002_a,Dietl:2007_b} Here, examples are Mn in ZnO and Fe in GaN,\cite{Pacuski:2007_a} which can trap a hole, despite that the corresponding donor states, Mn$^{2+}$/Mn$^{3+}$ and Fe$^{3+}$/Fe$^{4+}$, respectively reside in the valence band.

\subsection{Effects of exchange interaction}

In addition to forming trapping or doping centers, as discussed above, the magnetic impurities affect carrier localization via the strong $s$,$p$-$d$ exchange interaction between localized spins and effective mass carriers. This interaction gives rise to: (i) giant spin splitting of bands, which depends on the magnitude and direction of macroscopic magnetization of localized spins; (ii) spin-disorder scattering; (iii) formation of bound magnetic polarons (BMP); (iv) carrier-mediated ferromagnetic ordering. Owing to these phenomena, temperature and magnetic dependencies of conductivity in magnetic semiconductors differ dramatically from those known from studies of non-magnetic counterparts. Actually, since the synthesis of first magnetic semiconductors in the 1960s,\cite{Wachter:1979_a} diluted magnetic semiconductors (DMS) in the 1970s,\cite{Dietl:1994_a} and diluted ferromagnetic semiconductors in the 1990s,\cite{Matsukura:2002_c} the influence of localized spins on charge transport has been a central topic in the physics of these materials.

\subsection{Drude-Boltzmann effects}

Some of the phenomena specific to magnetically doped semiconductors can be interpreted by incorporating the above-mentioned exchange effects into the Drude-Boltzmann formalism. For instance, the giant magnetization-dependent spin splitting explains the behavior of Shubnikov-de Haas effect in DMS\cite{Jaczynski:1978_a} as well as the magnitudes of both anisotropic magnetoresistance\cite{Jungwirth:2006_a}  and  resistance shoulder at $T<T_{\mathrm{C}}$ in ferromagnetic semiconductors deeply in the metallic phase.\cite{Shapira:1974_a,Lopez-Sancho:2003_b} Furthermore, the a.~c. conductivity\cite{Jungwirth:2006_a,Hankiewicz:2004_c}  and  domain-wall resistance\cite{Chiba:2006_a} of (Ga,Mn)As can be understood in a semi-quantitative fashion within the Drude-Boltzmann-like approach.

Similarly, by adding the free energy of bound magnetic polarons to the acceptor binding energy, the temperature and field dependent conductance of \textit{p}-(Cd,Mn)Te in the thermal activation range has been described.\cite{Jaroszynski:1985_a} However, many experiments show that materials in question very often reside on or at the vicinity of MIT boundary.\cite{Dietl:1994_a,Matsukura:1998_a,Jungwirth:2007_a,Sheu:2007_a}  In such a case, the influence of spin phenomena upon the quantum  corrections to the Drude-Boltzmann conductivity determines the dependence of transport properties on the magnetic field and temperature.\cite{Dietl:1994_a}

\subsection{Quantum localization phenomena}
\subsubsection{Temperature dependent localization and colossal negative magnetoresistance}

Already early studies of magnetic semiconductors\cite{Wachter:1979_a} revealed that the exchange interaction between effective-mass carriers and disordered spins increases localization, {\em i.~e}, shifts MIT to higher carrier densities. Accordingly, a colossal decrease of the resistivity is observed when the spins of magnetic ions get aligned by the external magnetic field or when the transition to an ordered magnetic state takes place.

These finding are usually explained invoking the presence of either bound magnetic polarons (BMP) or a nanoscale phase separation.
According to the BMP scenario, put forward in the context of europium chalcogenides,\cite{Wachter:1979_a} the exchange interaction leads to the formation of a spin-polarized cloud of magnetic ions inside the Bohr sphere of occupied impurities. This increases the impurity binding energy, reduces $a_B$, and thus increases $n_c$. Within the nanoscale phase separation model, developed for colossal magnetoresistance oxides,\cite{Dagotto:2001_a} a competition between carrier-mediated ferromagnetism and intrinsic antiferromagnetism leads to a phase separation into magnetically ordered and disordered regions. This spatially inhomogeneous state sets in a temperature $T^*$ which is typically significantly higher than the Curie temperature $T_{\mathrm{C}}$ below which a {\em global} ferromagnetic order develops. Finally, it has been suggested\cite{Nagaev:2001_a} that carrier density fluctuations associated with the presence of ionized impurities lead to an enhancement of ionized-impurity scattering by the associated magnetization cloud, the reasoning known as the magnetic-impurity model.

According to data for \textit{n}-(Cd,Mn)Se collected in Fig.~4,\cite{Sawicki:1986_a,Dietl:1986_a,Glod:1994_b} a similar behavior has been found in DMS. In particular, the magnitude of resistance grows substantially on lowering temperature. This temperature-dependent localization observed also in \textit{n}-(Cd,Mn)Te,\cite{Terry:1992_a,Jaroszynski:2007_a}
\textit{n}-(Cd,Zn,Mn)Se,\cite{Smorchkova:1997_a} \textit{n}-(Zn,Mn)O,\cite{Andrearczyk:2005_a} and \textit{n}-(Zn,Co)O,\cite{Dietl:2007_d} can be removed by the magnetic field  resulting in colossal negative MR. Importantly, the effect exists only in the vicinity of MIT, which rules out the magnetic-impurity model.\cite{Nagaev:2001_a}

\begin{figure}[tb]
\begin{center}
\includegraphics[scale=0.37]{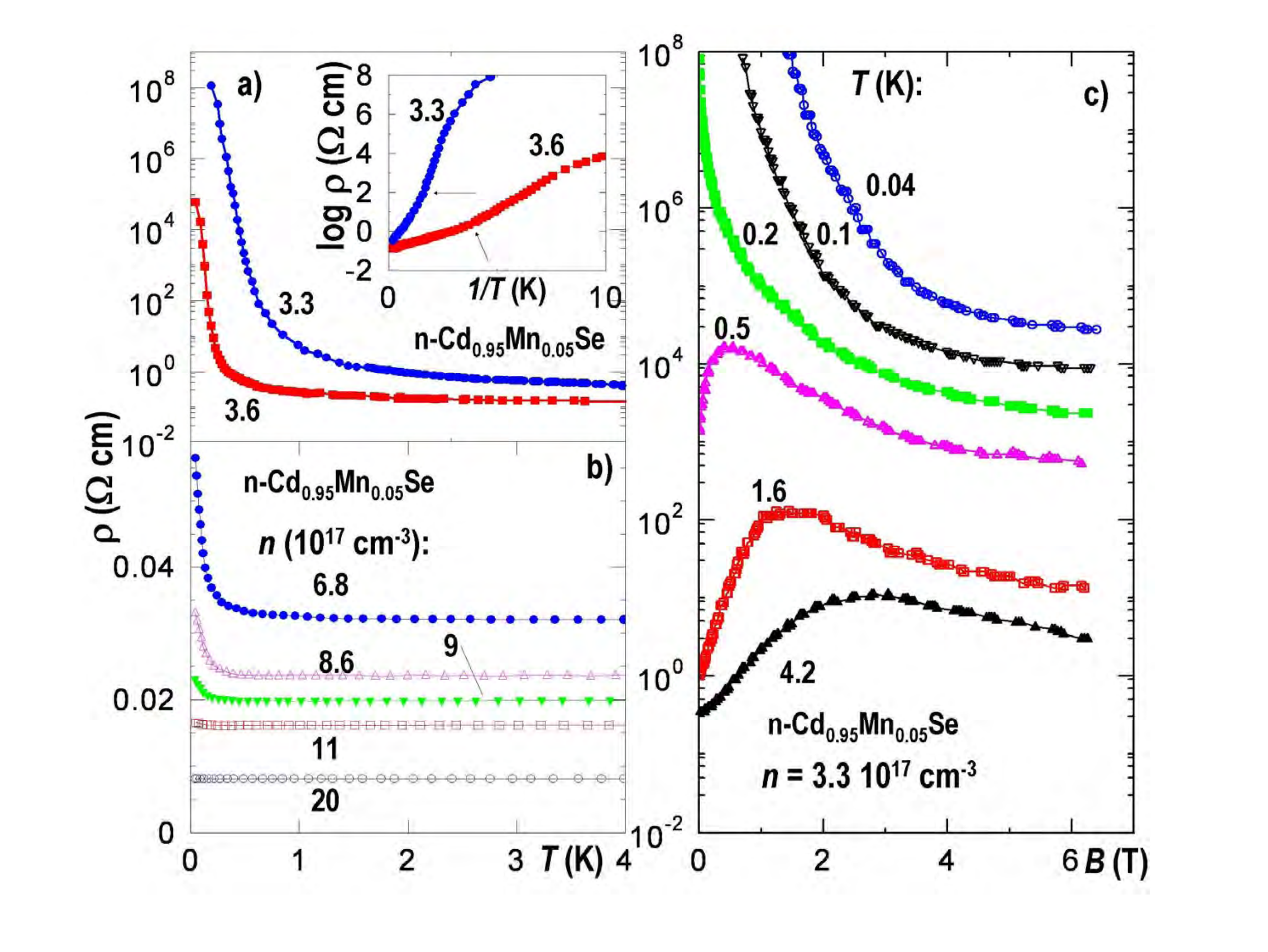}
\end{center}
\caption{(color online) Temperature dependence of resistivity in Cd$_{0.95}$Mn$_{0.05}$Se:In with various electron concentrations (left panels). Temperature dependent localization induced by the presence of magnetic ions is seen below $\sim 0.5$~K in samples sufficiently close to localization boundary. Magnetoresistance, consisting of the colossal negative component (originating from the destructive effect of the magnetic field on the temperature dependent localization) and the positive component (originating from the effect of the giant spin splitting on quantum corrections to conductivity), is shown in the right panel for one sample at various temperatures  (adapted after Sawicki  {\em et al.}\cite{Sawicki:1986_a, Dietl:1986_a,Glod:1994_b}).}
\label{f4}
\end{figure}

The above findings have been interpreted\cite{Sawicki:1986_a} in terms of the effect of spin-disorder scattering upon the quantum corrections to conductivity. Such scattering, if efficient enough, destroys quantitatively important "anti-localization" terms stemming from the disorder-modified carrier-carrier interactions. However, in order to explain the magnitude of the effect and its temperature dependence, the Mn spins have to form ferromagnetic bubbles whose magnetization should increase as magnetic susceptibility on lowering temperature.\cite{Sawicki:1986_a,Glod:1994_b} The formation of BMP around impurity-like states at the localization boundary could provide the require scattering centers.\cite{Sawicki:1986_a,Glod:1994_b} At the same time, thermodynamic fluctuations of magnetization cannot explain the observations as the corresponding scattering rate {\em decreases} with lowering temperature, $1/\tau_s \sim T\chi(T)$, where $\chi(T) \sim T^{-\alpha}$, where $ 0< \alpha < 1$.\cite{Glod:1994_b}

However, recent low-temperature MR studies of gated modulation-doped \textit{n}-(Cd,Mn)Te quantum wells revealed also the presence of temperature-dependent localization and associated colossal negative MR in the cross-over region between weak and strong localization, as shown in Fig.~5. This calls into question the BMP scenario, as not many impurity-like states are expected in such structures.
The proposed explanation\cite{Jaroszynski:2007_a} involves the model of nanoscale phase separation. In particular, low-temperature divergence of magnetic susceptibility, $\chi(T) \sim T^{-\alpha}$, points to the presence of the carrier-induced ferromagnetic instability  at $T \rightarrow 0$, which is hidden by short-range antiferromagnetic interactions, resulting in a non-magnetized ground state.\cite{Kechrakos:2005_a} These competing interactions, together with mesoscopic fluctuations in the local value of the density-of-states in the vicinity of  MIT, lead to the formation of randomly oriented and distributed mesoscopic ferromagnetic bubbles at $T^* > 0$. Because the bubbles can produce local spin splittings comparable to the Fermi energy, the conductance gets reduced. Furthermore, the bubbles constitute efficient killers of "anti-localization" terms referred to above and make the value of $n_c$ to increase on cooling and to decrease with the magnetic field.

\begin{figure}[tb]
\begin{center}
\includegraphics[scale=1.1]{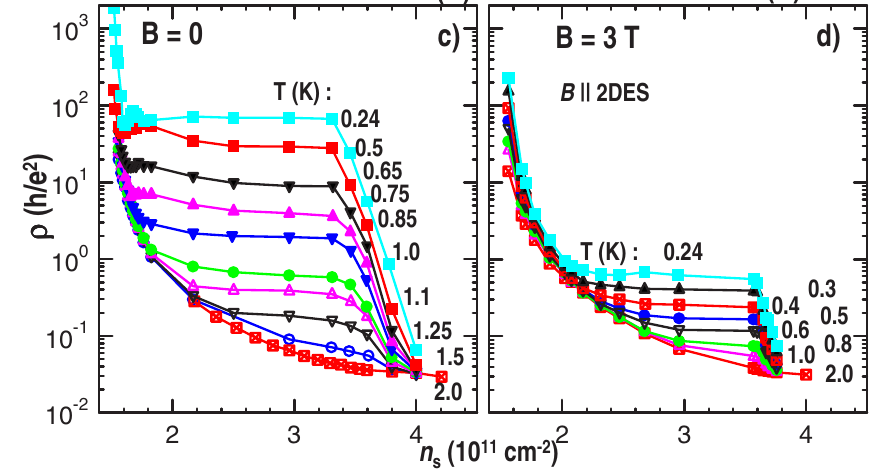}
\end{center}
\caption{(color online) Temperature dependence of resistivity in the absence of an external magnetic field (left panel) and in 3~T (right panel) for modulation-doped quantum well of Cd$_{0.985}$Mn$_{0.015}$Te with various sheet densities changed by the gate voltage. The data give evidence for temperature dependent localization and colossal negative magnetoresistance (after Jaroszynski  {\em et al.}\cite{Jaroszynski:2007_a} reproduce with the permission, Copyright (2007) by the American Physical Society).}
\label{f5}
\end{figure}

As holes mediate more effectively ferromagnetic correlations, leading to $T_{\mathrm{C}} \gtrsim 2$~K \textit{p}-type (Zn,Mn)Te,\cite{Ferrand:2001_a} the relevant temperature scale $T^*$ is shifted towards  higher temperatures in \textit{p}-type systems. Furthermore, as expected within this model, for samples on the insulator side of MIT, even at $T \ll T_{\mathrm{C}}$ only a fraction of spins is aligned ferromagnetically.\cite{Ferrand:2001_a} It has to be noted that at this stage there is no quantitative theory describing temperature dependent localization in DMS.

\subsubsection{Spin-splitting induced negative magnetoresistance}

As mentioned above, the field-induced ordering of localized spins leads to colossal negative MR. This MR can be enhanced further on by spin-splitting-induced redistribution of carriers between spin subbands, which shifts the Fermi energy away from the localization boundary.\cite{Fukuyama:1979_a} The effect is particularly strong in \textit{p}-type DMS, where the redistribution increases the participation of light holes in charge transport.  Actually this effect accounts mainly for the field-induced insulator-to-metal transition in \textit{p}-(Hg,Mn)Te presented in Figs.~2 and 3.

\subsubsection{Spin-splitting-induced positive magnetoresistance}

The "anti-localization" terms whose destruction by spin-disorder scattering leads to temperature dependent localization, can also be partly destroyed by spin splitting. Due to the giant spin splitting, the effect is particularly strong in DMS, where it scales rather with magnetization than with the magnetic field.  As shown in Figs.~4 and 6, where experimental findings for \textit{n}-(Cd,Mn)Se and \textit{n}-(Zn,Mn)O are presented,\cite{Andrearczyk:2005_a}  spin-splitting-induced positive MR competes at high temperatures with negative MR caused by the orbital weak-localization effect, whereas at low temperatures it is masked by temperature dependent localization and the associated colossal negative MR. As seen, weak-localization theory\cite{Altshuler:1985_a,Fukuyama:1985_a,Lee:1985_a} describes satisfactorily the data, except for the low temperature regime, as no quantitative model is presently available for temperature dependent localization.

\begin{figure}[tb]
\begin{center}
\includegraphics[scale=0.8]{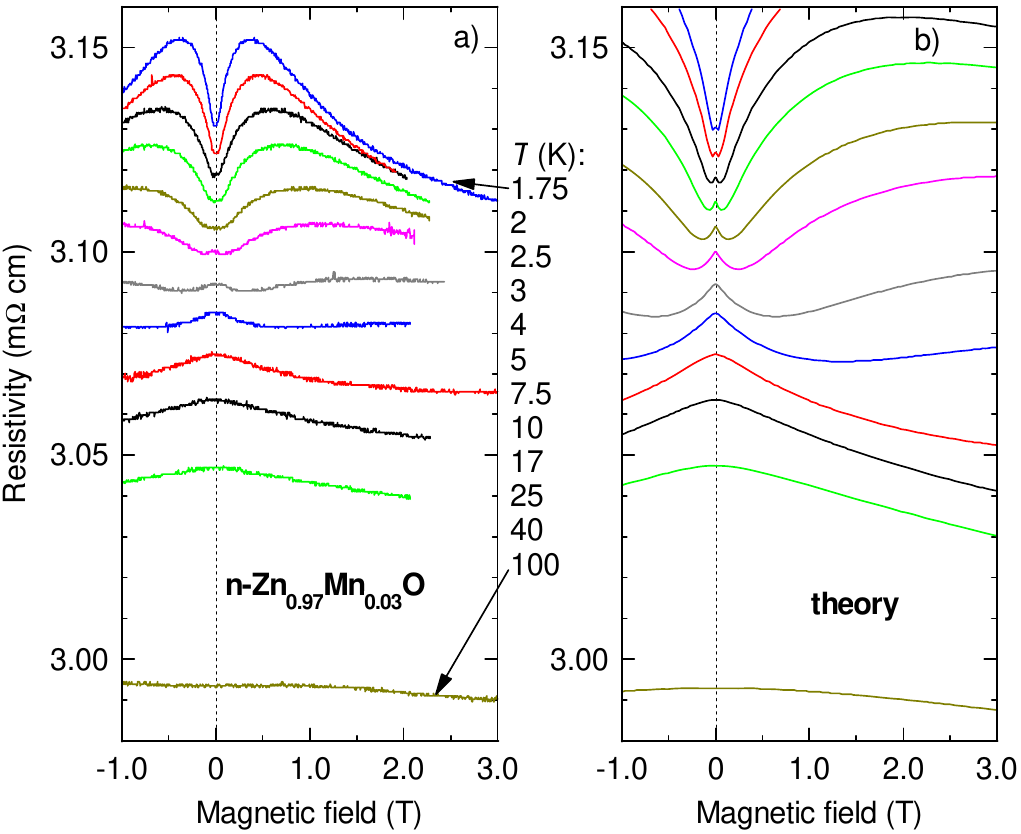}
\end{center}
\caption{(color online) Experimental (a) and calculated with no adjustable
parameter (b) resistivity changes in the magnetic field for
n-Zn$_{0.97}$Mn$_{0.03}$O. Low temperature negative magnetoresistance is not described by theory as it does not take into account temperature dependent localization (after Andrearczyk  {\em et al.}\cite{Andrearczyk:2005_a} reproduce with the permission, Copyright (2005) by the American Physical society).}
\label{f6}
\end{figure}

A similar behavior, {\em i.~e.},
the appearance of temperature-dependent {\em positive} MR in the
presence of magnetic ions, has been observed also for
\textit{n}-(Cd,Mn)Se,\cite{Sawicki:1986_a}
\textit{n}-(Cd,Mn)Te,\cite{Shapira:1990_a,Jaroszynski:2007_a}
\textit{n}-(Cd,Zn,Mn)Se,\cite{Smorchkova:1997_a} and \textit{n}-(Zn,Co)O,\cite{Dietl:2007_d} and
quantitatively described, as in Fig.~6, by the effect of the field-induced giant spin splitting on disorder-modified electron-electron interactions. In accord with such an interpretation, no corresponding MR was found in ferromagnetic (Ga,Mn)As,\cite{Matsukura:2004_b} where hole
states are spin-polarized already in the absence of an external
magnetic field.

\subsubsection{Effects of spin splitting on Hall resistance}

The quantum localization phenomena affect also the behavior of Hall conductivity in disordered systems.\cite{Altshuler:1985_a,Fukuyama:1985_a,Lee:1985_a,Dugaev:2001_a} Actually, sizable anomalies in the Hall resistance of II-VI \textit{n}-type DMS have been observed.\cite{Sawicki:1988_a,Cumings:2006_a} They have been assigned to the influence of the giant spin splitting on quantum localization\cite{Sawicki:1988_a} or on the anomalous Hall effect.\cite{Cumings:2006_a}

\subsection{Carrier-mediated ferromagnetism at localization boundary}

The Curie temperature $T_{\mathrm{C}}$ of (Ga,Mn)As\cite{Matsukura:1998_a} and p-(Zn,Mn)Te\cite{Ferrand:2001_a} one of the thermodynamic characteristics, shows no critical behavior at MIT. At the same time, $T_{\mathrm{C}}$ of (Ga,Mn)As vanishes rather rapidly when moving away from MIT into the insulator phase,\cite{Matsukura:1998_a,Sheu:2007_a} while it grows steadily with the magnitude of the conductivity on the metal side of MIT.\cite{Matsukura:1998_a,Potashnik:2001_a,Campion:2003_b} Guided by these observations, the band scenario has been proposed in order to describe the ferromagnetism in ferromagnetic III-V and II-VI semiconductors on both sides of MIT.\cite{Dietl:2000_a,Ferrand:2001_a,Dietl:2001_b}

Within this model, the hole localization length, which diverges at MIT, remains much greater than the average distance between acceptors for the experimentally important range of hole densities.  Thus, holes can be regarded as band-like at the length scale relevant for the coupling between magnetic ions. Hence, the spin-spin exchange interactions are effectively mediated by the itinerant carriers, so that the $p$-$d$ Zener model can be applied also to the insulator side of MIT. This view has been strongly supported by results of inelastic neuron scattering of nearest neighbor Mn pairs in \textit{p}-(Zn,Mn)Te.\cite{Kepa:2003_a} In this experiment, the hole-induced change in the pair interaction energy shows the value expected for the band carriers despite that the studied sample was on the insulator side of MIT.

Since large mesoscopic spatial fluctuations in the magnitude of the density-of-states are expected near MIT, the ferromagnetic order develops locally already at $T^* > T_{\mathrm{C}}$.\cite{Mayr:2002_a}  This disorder-induced cluster ferromagnetism explains temperature dependent localization, as discussed above, as well elucidates why in samples on the insulator side of MIT only a fraction of spins is aligned ferromagnetically.\cite{Oiwa:1997_a,Ferrand:2001_a,Sheu:2007_a}

\section{Quantum localization effects in (Ga,Mn)As}

\subsection{Metal-insulator transition and onset of ferromagnetism in (Ga,Mn)As}

Comparing to GaAs doped with shallow acceptors such as Carbon, the Mn impurity introduces a stronger local potential stemming from  sizable $p$-$d$ hybridization. This increases the Mn acceptor ionization energy $E_I$, reduces $a_B$ and, thus, enlarges $n_c$, perhaps by one order of magnitude, comparing to, {\em e.~g.}, GaAs:C. This shift of $n_c$ is presumably smaller in InSb:Mn and InAs:Mn but even larger in GaP:Mn and GaN:Mn where $E_I \approx 1$~eV due to the short bond length and strong $p$-$d$ hybridization.\cite{Dietl:2007_b} Within the model put forward here, once MIT is approached from the insulator side, many-body screening washes out bound states and makes the appearance of itinerant holes capable of transmitting magnetic interactions between diluted spins. The band scenario can serve for the description of ferromagnetism in this regime.

Experimentally,\cite{Sheu:2007_a} the onset of the carrier-mediated ferromagnetism is located rather close to MIT in (Ga,Mn)As, in agreement with the notion that the presence of band-like states are necessary carrier-mediated ferromagnetism. Comparing to \textit{p}-(Zn,Mn)Te, where $T_{\mathrm{C}}$  exceeds barely 5~K, $T_{\mathrm{C}}$ in (Ga,Mn)As reaches rather fast a 20--30 K level. This difference is associated with a higher value of the density-of-states at $n_c$ and the absence of competing antiferromagnetic interactions in (Ga,Mn)As.\cite{Dietl:2001_b}
Actually, ferromagnetic order starts to develop at $T^* > T_{\mathrm{C}}$ in the regions where the local carrier density is large enough to support long-range ferromagnetic correlations between randomly distributed Mn spins.\cite{Mayr:2002_a}

As could be expected for samples on the insulator side of MIT, in which the disorder-driven fluctuations of the local density-of-states are particularly large, the field\cite{Oiwa:1997_a} and temperature\cite{Sheu:2007_a} dependence of magnetization shows that even at $T \ll T_{\mathrm{C}}$ only a fraction of spins is aligned ferromagnetically. According to this model, the portion of the material encompassing the ferromagnetic bubbles, and thus the magnitude of the spontaneous ferromagnetic moment, grows with the net acceptor concentration, extending over the whole sample volume well within the metal phase.\cite{Jungwirth:2006_b}

\subsection{Temperature dependence of resistance in (Ga,Mn)As}

As argued above, a characteristic feature of carrier-controlled ferromagnetic semiconductors at the localization boundary is the presence of randomly oriented ferromagnetic bubbles, which start to develop at $T^* >  T_{\mathrm{C}}$. Together with critical thermodynamic fluctuations that develop  $T \rightarrow  T^+_{\mathrm{C}}$, they  account for a resistance maximum near $ T_{\mathrm{C}}$ and the associated negative magnetoresistance, as shown in Fig.~7.\cite{Matsukura:1998_a} Both effects disappear deeply in the metallic and isolating phases. The underlying physics is analogous to that accounting for similar anomalies, albeit at much lower temperatures, in $n$-type DMS at the localization boundary, as discussed above. According to this insight, the presence of randomly oriented ferromagnetic bubbles has two consequences. First they constitute potential barriers of the order of the Fermi energy, which diminish conductance. Second, they give rise to efficient spin-disorder scattering of the carriers. Such scattering reduce "anti-localization" corrections to conductivity near the Anderson-Mott transition.\cite{Belitz:1994_a,Altshuler:1985_a,Fukuyama:1985_a,Lee:1985_a} In particular, spin-disorder scattering, once more efficient than spin-orbit scattering, destroys "anti-localization" quantum corrections to the conductivity associated with the one-particle interference effect\cite{Timm:2005_a} and with the particle-hole triplet channel.\cite{Jaroszynski:2007_a} While both phenomena increase the resistance value upon approaching $T_{\mathrm{C}}$, the latter -- resulting from disorder-modified carrier correlation -- is usually quantitatively more significant.\cite{Belitz:1994_a,Altshuler:1985_a,Fukuyama:1985_a,Lee:1985_a}

\begin{figure}[tb]
\begin{center}
\includegraphics[scale=1.05]{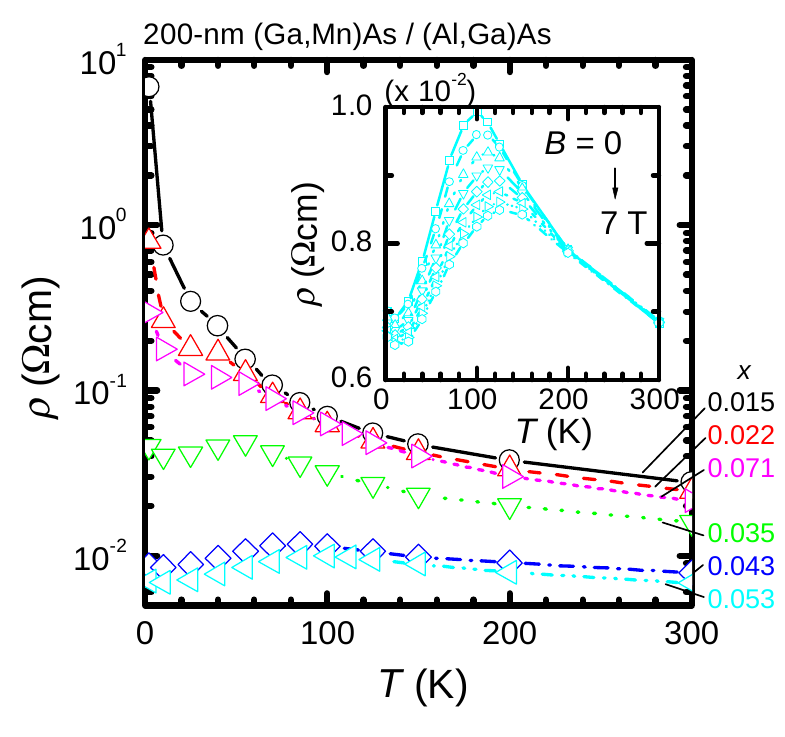}
\end{center}
\caption{(color online) Temperature dependence of resisitivity for Ga$_{1-x}$Mn$_x$As films on the both sides of the metal-insulator transition. Resistance maximum in the vicinity of the Curie temperature and the associated negative magnetoresistance (inset) are seen (after Matsukura {\em et al.}\cite{Matsukura:1998_a} reproduce with the permission, Copyright (1998) by the American Physical Society).}
\label{f7}
\end{figure}

As expected within the above model, the resistance maximum tends to disappear deeply in the insulator phase\cite{Sheu:2007_a} (where ferromagnetic bubbles occupy only a small portion of the sample) as well as deeply in the metallic phase\cite{Potashnik:2001_a} (where ferromagnetic alignment is uniform and quantum localization unimportant). In the metallic region, the resistance is approximately constant down to $T_{\mathrm{C}}$ and gradually decreases at lower temperatures. The Drude-Boltzmann approach taking into account the spin-splitting-induced carrier redistribution between spin subbands describes satisfactorily the data in this range.\cite{Lopez-Sancho:2003_b} Furthermore, the disappearance of the resistance maximum away from MIT appears to suggest that contributions expected from critical scattering\cite{Omiya:2000_a} and magnetic-impurity\cite{Nagaev:2001_a,Yuldashev:2003_a,Goennenwein:2005_a} models are quantitatively unimportant, a conclusion indirectly supported by the fact that rather large values of the exchange integral $\beta$ had to be assumed to fit the temperature dependence of resistance near  $T_{\mathrm{C}}$ in (Ga,Mn)As samples close to MIT within those models.\cite{Omiya:2000_a,Yuldashev:2003_a}

In addition to the resistance changes near $T_{\mathrm{C}}$, a sizable resistance increase with lowering temperature shows up at $T \ll T_{\mathrm{C}}$ in metallic (Ga,Mn)As, an onset of the effect visible in Fig.~7. It is known that carrier spin polarization destroys the Kondo effect. Actually,  this upturn of resistance can be explained in terms of quantum corrections to the conductivity in the weakly localized regime\cite{Matsukura:2004_b} for the spin-polarized universality class.\cite{Wojtowicz:1986_a} Assuming that only singlet electron-hole channel is relevant in this case, the temperature dependence of conductance is given by\cite{Altshuler:1985_a,Fukuyama:1985_a,Lee:1985_a}
\begin{equation}
\Delta \sigma = \frac{0.915e^2}{3\pi^2\hbar}\sqrt{\frac{k_BT}{\hbar D}}
\end{equation}
 or
\begin{equation}
\Delta\sigma = 4.4\sqrt{\frac{m^*T\left[\mbox{K}\right]}{m_ok_Fl}}
\left[\Omega\mbox{cm}\right]^{-1}.
\end{equation}
We see that since $m^*/(m_ok_Fl) \approx 1$, the expected change of conductance between 1 and 4~K is about 4.4~($\Omega$cm)$^{-1}$, in a good agreement with the experimental data summarized in Ref.~\citeonline{Honolka:2007_a}. This reconfirms the band character of the relevant states in (Ga,Mn)As. Importantly, the proximity to MIT\cite{Wojtowicz:1986_a,Honolka:2007_a} or the dimensional cross-over in thin films\cite{Neumaier:2007_a} will modify the $T^{1/2}$ dependence at low temperatures.

\subsection{Magnetoresistance of (Ga,Mn)As}

It is important to begin by noting that an additional flavor of \textit{p}-type ferromagnetic semiconductors is a large magnitude of the anomalous Hall effect (AHE). This should be taken into account when determining the magnitudes of conductivity tensor components from resistivity measurements, even in the absence of an external magnetic field at $T< T_{\mathrm{C}}$.

It is convenient to discuss MR separately in four regions. In the temperature range near $T_{\mathrm{C}}$, the magnetic-field orientation of preformed magnetic bubbles as well as the rapid polarization of the Mn spins work together to remove the enhancement of resistance.

However, the negative MR persists in the field and temperature region, where Mn spins are entirely polarized. As shown in Fig.~8, this high-field MR can be quantitatively describes in terms of the weak-localization orbital effect,\cite{Matsukura:2004_b} which near MIT shows up in the cyclotron energy is greater than the spin relaxation rates.\cite{Altshuler:1985_a,Fukuyama:1985_a,Lee:1985_a,Dugaev:2001_b} It worth recalling at this point that the spin splitting specific to the ferromagnetic phase reduces spin-scattering rates rather substantially. Interestingly, the orbital effect in question accounts presumably for the insulator-to-metal transition revealed for (Ga,Mn)As in high magnetic fields.\cite{Katsumoto:1998_a}

\begin{figure}[tb]
\begin{center}
\includegraphics[scale=1]{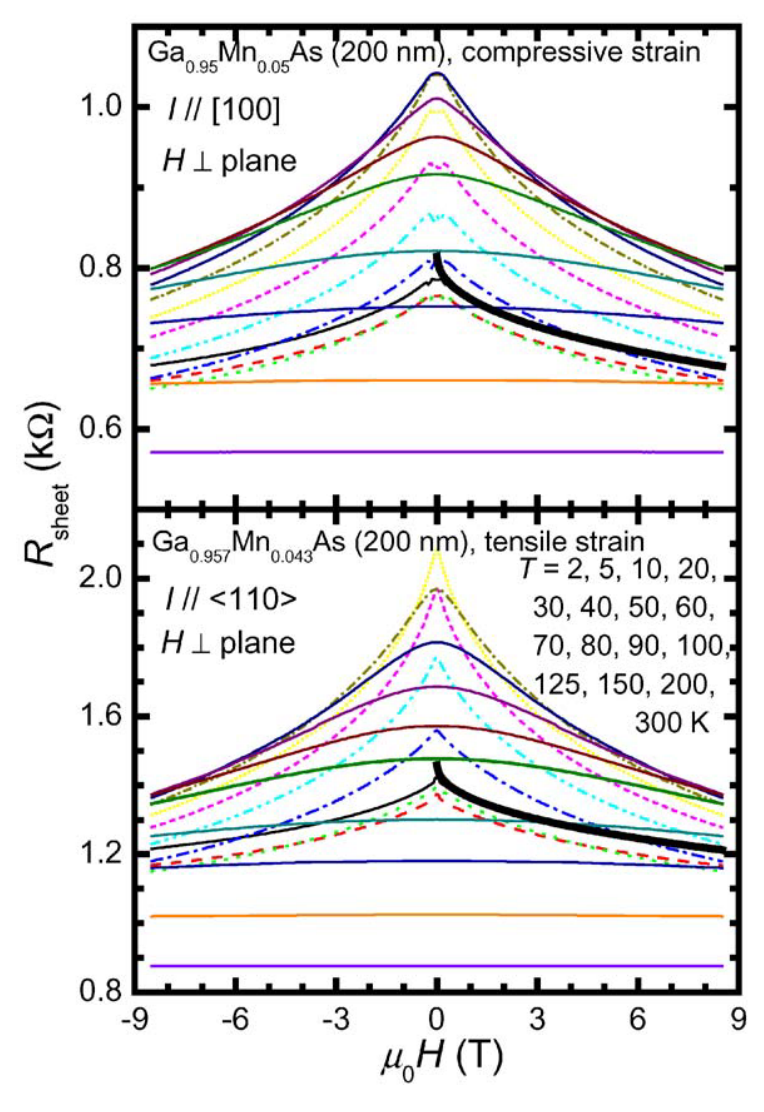}
\end{center}
\caption{(color online) Field and temperature dependencies of resistance in
Ga$_{0.95}$Mn$_{0.05}$As on GaAs (compressive strain, upper panel) and in tensile strained Ga$_{0.957}$Mn$_{0.043}$As on (In,Ga)As (lower panel) for
magnetic field perpendicular to the film plane. Starting from up,
subsequent curves at $H = 0$ correspond to temperatures in K: 70,
60, 80, 50, 90, 40, 100, 30, 125, 20, 2, 5, 10, 150, 200, 300 (upper panel) and to 50, 60, 40, 70, 30, 80, 90, 20, 100, 2, 10, 5,
125, 150, 200, 300 (lower panel). The thick solid lines on 2~K
data in positive magnetic field side show the fitting to weak localization theory, indicating that negative magnetoresistance originates from the effect of the magnetic field on interferences of scattering amplitude (after Matsukura {\em et al.}\cite{Matsukura:2004_b} reproduce with the permission, Copyright (2004) by Elsevier).}
\label{f8}
\end{figure}

The third region of interest corresponds to the magnetic fields determined by the magnitude of the anisotropy energy. In this regime, effects related to the field-induced reorientation of spontaneous magnetization dominate. As known, in the presence of spin-orbit interaction, non-zero values of magnetization and/or strain lead to a specific anisotropy of the conductivity tensor components already within the Drude-Boltzmann approach. This results in a dependence of the resistance magnitude on the direction of magnetization\cite{Jungwirth:2006_a} as well as to the appearance of the so-called planar Hall effect.\cite{Tang:2003_a} If all hole subbands remain occupied, anisotropic magnetoresistance (AMR) is of the order of a few percent.\cite{Jungwirth:2006_a,Matsukura:2004_b} However, when the Fermi energy becomes smaller than the subband spin splitting, AMR and the related tunneling anisotropic magnetoresistance (TAMR) can be quite large,\cite{Sankowski:2007_a} especially near MIT.\cite{Pappert:2006_a} According to the results presented in Fig.~2,\cite{Wojtowicz:1986_a} despite of a large value of AMR, the critical point remains invariant to the relative orientation of the current and magnetization direction in cubic (Hg,Mn)Te. This may not be, however, the case of strained (Ga,Mn)As films, where the critical point can be displaced by changing the direction of magnetization.\cite{Pappert:2006_a}

Finally, we turn to possible origins of a linear MR across zero magnetic field and of an associated resistance jump at the coercive field $H_c$, which have been detected experimentally in ferromagnetic (Ga,Mn)As.\cite{Goennenwein:2005_a,Chiba:2006_a} As discussed above, spin splitting leads to sizable positive and negative contributions to MR in \textit{n}-type and \textit{p}-type II-VI DMS, respectively. Furthermore, AMR and AHE, rather sizable in (Ga,Mn)As, originate also from a non-zero value of spin splitting. Obviously, at $T \ll T_{\mathrm{C}}$, the spins are entirely polarized so that the giant exchange spin splitting does not vary with the magnitude of the external magnetic field. However, it may depend on the field orientation in respect to the crystallographic directions.  Furthermore, the total spin splitting consists of the giant exchange contribution and a smaller anisotropic band term linear in the magnetic field. Similarly, the Hall conductance contains also a normal term that is linear in the field. Another mechanism is possible if a part of localized spins is coupled by an antiferromagnetic interaction or not coupled at all. In such a case spin-disorder scattering rate  contains a term linear in the field-induced magnetization.\cite{Csontos:2005_a} Which of these effects dominates is unclear by now but since the relative sign of the field-independent and field-dependent terms changes at $H_c$, both linear MR and resistance jump at $H_c$ are expected for the mechanisms discussed here, as observed.\cite{Goennenwein:2005_a,Chiba:2006_a,Csontos:2005_a}

\section{Summary}

The results discussed in this paper reemphasize the significance of quantum localization effects in the physics of diluted magnetic and ferromagnetic semiconductors. These phenomena originate from interference of scattered waves and interference of carrier-carrier interaction amplitudes. These two interference manifestations have to be considered on equal footing and they account for a rather complex field, magnetization, and temperature dependence of resistance. In a wide region around the metal-insulator critical point, the quantum phenomena coexist with or dominate over Drude-Bolzmann effects, such as anisotropic magnetoresistance.

The insight gained from studies of II-VI Mn-based DMS has been exploited here to describe magnetotransport phenomena in (Ga,Mn)As and related compounds. According to the present understanding, the high-field positive magnetoconductance of (Ga,Mn)As, $\Delta\sigma \sim B^{1/2}$, present in virtually all other doped semiconductors, results from the orbital weak-localization effect. The resistance maximum and the associated negative magnetoresistance near the Curie temperature are assigned to the cluster ferromagnetism specific to magnetically doped semiconductors at the localization boundary. The randomly oriented ferromagnetic bubbles reduce the conductance directly as well as by destroying the "anti-localization" quantum corrections driven by disorder modified carrier-carrier interactions. The  randomness-driven mesoscopic magnetization fluctuations lead also to a significant increase of low-temperature resistance in \textit{n}-type II-VI Mn-based DMS in which $T_{\mathrm{C}}$ is rather low. In ferromagnetic \textit{p}-(Ga,Mn)As, where the Mn spins are entirely polarized at low temperatures, the interaction effects result in a standard square root increase of conductance, $\Delta\sigma \sim T^{1/2}$ at $T \ll  T_{\mathrm{C}}$, which can be modified by the dimensional cross-over in thin films.  Furthermore, the sensitivity of conductance to spin splitting may explain resistance anomalies at coercive fields, where the relative direction of the molecular and external magnetic fields changes.

In this paper, we have assumed that the localized spins are distributed randomly. Nevertheless, as we have argued,  non-uniformities of magnetization occur due to large fluctuations in the local density of carrier states near MIT. However, in a number of DMS, the distribution of magnetic ions is highly non-random. Actually, nano-scale regions containing a large concentration of the magnetic constituent appear to account for the high-temperature ferromagnetism observed in a variety of such systems. This aspect of DMS has recently been reviewed elsewhere.\cite{Dietl:2007_a,Dietl:2007_e}

\section*{Acknowledgments}
This work was supported in part by NANOSPIN E.~C. project (FP6-2002-IST-015728), by the Humboldt Foundation, and carried out in collaboration with T. Andrearczyk, P. G{\l}\'od, M. Sawicki, and T. Wojtowicz in Warsaw, J. Jaroszy\'nski in Tallahassee,  F. Matsukura and H. Ohno in Sendai, and with G. Bauer and A. Bonanni in Linz.


\begin{thebibliography}{99}
\expandafter\ifx\csname natexlab\endcsname\relax\def\natexlab#1{#1}\fi
\expandafter\ifx\csname bibnamefont\endcsname\relax
  \def\bibnamefont#1{#1}\fi
\expandafter\ifx\csname bibfnamefont\endcsname\relax
  \def\bibfnamefont#1{#1}\fi
\expandafter\ifx\csname citenamefont\endcsname\relax
  \def\citenamefont#1{#1}\fi
\expandafter\ifx\csname url\endcsname\relax
  \def\url#1{\texttt{#1}}\fi
\expandafter\ifx\csname urlprefix\endcsname\relax\def\urlprefix{URL }\fi
\providecommand{\bibinfo}[2]{#2}
\providecommand{\eprint}[2][]{\url{#2}}

\bibitem{Matsukura:2002_c}
\bibinfo{author}{\bibfnamefont{F.}~\bibnamefont{Matsukura}},
  \bibinfo{author}{\bibfnamefont{H.}~\bibnamefont{Ohno}} \bibnamefont{and}
  \bibinfo{author}{\bibfnamefont{T.}~\bibnamefont{Dietl}}, in
  \emph{\bibinfo{booktitle}{Handbook of Magnetic Materials}}, edited by
  \bibinfo{editor}{\bibfnamefont{K.}~\bibnamefont{Buschow}}
  (\bibinfo{publisher}{Elsevier, Amsterdam}, \bibinfo{year}{2002}),
  vol.~\bibinfo{volume}{14}, p.~\bibinfo{pages}{1}.

\bibitem{Jungwirth:2006_a}
\bibinfo{author}{\bibfnamefont{T.}~\bibnamefont{Jungwirth}},
  \bibinfo{author}{\bibfnamefont{J.}~\bibnamefont{Sinova}},
  \bibinfo{author}{\bibfnamefont{J.}~\bibnamefont{{Ma\v{s}ek}}},
  \bibinfo{author}{\bibfnamefont{J.}~\bibnamefont{{Ku\v{c}era}}}
  \bibnamefont{and} \bibinfo{author}{\bibfnamefont{A.~H.}
  \bibnamefont{MacDonald}}, \bibinfo{journal}{Rev. Mod. Phys.}
  \textbf{\bibinfo{volume}{78}}, \bibinfo{pages}{809}
  (\bibinfo{year}{2006}).

\bibitem{Dietl:2007_c}
\bibinfo{author}{\bibfnamefont{T.}~\bibnamefont{Dietl}},
  \bibinfo{author}{\bibfnamefont{H.}~\bibnamefont{Ohno}} \bibnamefont{and}
  \bibinfo{author}{\bibfnamefont{F.}~\bibnamefont{Matsukura}},
  \bibinfo{journal}{IEEE-Trans. Electronic Devices}
  \textbf{\bibinfo{volume}{54}}, \bibinfo{pages}{945}
  (\bibinfo{year}{2007}).

\bibitem{Belitz:1994_a}
\bibinfo{author}{\bibfnamefont{D.}~\bibnamefont{Belitz}} \bibnamefont{and}
  \bibinfo{author}{\bibfnamefont{T.~R.} \bibnamefont{Kirkpatrick}},
  \bibinfo{journal}{Rev. Mod. Phys.} \textbf{\bibinfo{volume}{66}},
  \bibinfo{pages}{261} (\bibinfo{year}{1994}).

\bibitem{Shklovskii:1984_a}
\bibinfo{author}{\bibfnamefont{B.~I.} \bibnamefont{Shklovskii}}
  \bibnamefont{and} \bibinfo{author}{\bibfnamefont{A.~L.} \bibnamefont{Efros}},
  \emph{\bibinfo{title}{Electronic Properties of Doped Semiconductors}}
  (\bibinfo{publisher}{Springer, Berlin}, \bibinfo{year}{1984}).

\bibitem{Palankovski:1999_a}
\bibinfo{author}{\bibfnamefont{V.}~\bibnamefont{Palankovski}},
  \bibinfo{author}{\bibfnamefont{G.}~\bibnamefont{Kaiblinger-Grujin}}
  \bibnamefont{and}
  \bibinfo{author}{\bibfnamefont{S.}~\bibnamefont{Selberherr}},
  \bibinfo{journal}{Materials Sci. Engin. B} \textbf{\bibinfo{volume}{66}},
  \bibinfo{pages}{46} (\bibinfo{year}{1999}).

\bibitem{Jungwirth:2007_a}
  \bibinfo{author}\bibnamefont{T.~Jungwirth}, \bibinfo{author}\bibfnamefont{Jairo~Sinova},
  \bibinfo{author}{\bibfnamefont{A.~H.} \bibnamefont{MacDonald}},
 \bibinfo{author}{\bibfnamefont{B.~L.} \bibnamefont{Gallagher}},
 \bibinfo{author}{\bibfnamefont{V} \bibnamefont{Nov\'ak}},
 \bibinfo{author}{\bibfnamefont{K.~V.} \bibnamefont{Edmonds}},
  \bibinfo{author}{\bibfnamefont{A.~W.} \bibnamefont{Rushforth}},
 \bibinfo{author}{\bibfnamefont{R.~P.} \bibnamefont{Campion}},
 \bibinfo{author}{\bibfnamefont{C.~T.} \bibnamefont{Foxon}},
  \bibinfo{author}{\bibfnamefont{L.}~\bibnamefont{Eaves}},
  \bibinfo{author}{\bibfnamefont{E.}~\bibnamefont{Olejník}},
  \bibinfo{author}{\bibfnamefont{J.}~\bibnamefont{{Ma\v{s}ek}}},
  \bibinfo{author}{\bibfnamefont{S.-R.~E.} \bibnamefont{Yang}},
  \bibinfo{author}{\bibfnamefont{J.}~\bibnamefont{Wunderlich}},
  \bibinfo{author}{\bibfnamefont{C.}~\bibnamefont{Gould}},
  \bibinfo{author}{\bibfnamefont{L.~W.} \bibnamefont{Molenkamp}},
 \bibinfo{author}{\bibfnamefont{T. Dietl} \bibnamefont{Molenkamp}}
  \bibnamefont{and}
   \bibinfo{author}{\bibfnamefont{H. Ohno} \bibnamefont{Ohno}}
  \bibinfo{journal}{Phys. Rev. B)}
  \textbf{\bibinfo{volume}{76}}, \bibinfo{pages}{125206}
  (\bibinfo{year}{2007}).

\bibitem{Ohno:2007_a}
\bibinfo{author}{\bibfnamefont{H.}~\bibnamefont{Ohno}} \bibnamefont{and}
  \bibinfo{author}{\bibfnamefont{T.}~\bibnamefont{Dietl}}, \bibinfo{journal}{J.
  Magn. Magn. Mat.}  (\bibinfo{year}{in press}).

\bibitem{Liu:1993_a}
\bibinfo{author}{\bibfnamefont{S.}~\bibnamefont{Liu}},
  \bibinfo{author}{\bibfnamefont{K.}~\bibnamefont{Karrai}},
  \bibinfo{author}{\bibfnamefont{F.}~\bibnamefont{Dunmore}},
  \bibinfo{author}{\bibfnamefont{H.~D.} \bibnamefont{Drew}},
  \bibinfo{author}{\bibfnamefont{R.}~\bibnamefont{Wilson}} \bibnamefont{and}
  \bibinfo{author}{\bibfnamefont{G.~A.} \bibnamefont{Thomas}},
  \bibinfo{journal}{Phys. Rev. B} \textbf{\bibinfo{volume}{48}},
  \bibinfo{pages}{11394} (1993).

\bibitem{Rosenbaum:1990_a}
\bibinfo{author}{\bibfnamefont{T.~F.} \bibnamefont{Rosenbaum}},
  \bibinfo{author}{\bibfnamefont{S.}~\bibnamefont{Pepke}},
  \bibinfo{author}{\bibfnamefont{R.~N.} \bibnamefont{Bhatt}} \bibnamefont{and}
  \bibinfo{author}{\bibfnamefont{T.~V.} \bibnamefont{Ramakrishnan}},
  \bibinfo{journal}{Phys. Rev. B} \textbf{\bibinfo{volume}{42}},
  \bibinfo{pages}{11214} (\bibinfo{year}{1990}).

\bibitem{Paalanen:1988_a}
\bibinfo{author}{\bibfnamefont{M.~A.} \bibnamefont{Paalanen}},
  \bibinfo{author}{\bibfnamefont{J.~E.} \bibnamefont{Graebner}},
  \bibinfo{author}{\bibfnamefont{R.~N.} \bibnamefont{Bhatt}} \bibnamefont{and}
  \bibinfo{author}{\bibfnamefont{S.}~\bibnamefont{Sachdev}},
  \bibinfo{journal}{Phys. Rev. Lett.} \textbf{\bibinfo{volume}{61}},
  \bibinfo{pages}{597} (\bibinfo{year}{1988}).

\bibitem{Altshuler:1985_a}
\bibinfo{author}{\bibfnamefont{B.~L.} \bibnamefont{Altshuler}}
  \bibnamefont{and} \bibinfo{author}{\bibfnamefont{A.~G.}
  \bibnamefont{Aronov}}, in \emph{\bibinfo{booktitle}{Electron-Electron
  Interactions in Disordered Systems}}, edited by
  \bibinfo{editor}{\bibfnamefont{A.~L.} \bibnamefont{Efros}} \bibnamefont{and}
  \bibinfo{editor}{\bibfnamefont{M.}~\bibnamefont{Pollak}}
  (\bibinfo{publisher}{North Holland, Amsterdam}, \bibinfo{year}{1985}),
  p.~\bibinfo{pages}{1}.

\bibitem{Fukuyama:1985_a}
\bibinfo{author}{\bibfnamefont{H.}~\bibnamefont{Fukuyama}}, in
  \emph{\bibinfo{booktitle}{Electron-Electron Interactions in Disordered
  Systems}}, edited by \bibinfo{editor}{\bibfnamefont{A.~L.}
  \bibnamefont{Efros}} \bibnamefont{and}
  \bibinfo{editor}{\bibfnamefont{M.}~\bibnamefont{Pollak}}
  (\bibinfo{publisher}{North Holland, Amsterdam}, \bibinfo{year}{1985}), p.
  \bibinfo{pages}{155}.

\bibitem{Lee:1985_a}
\bibinfo{author}{\bibfnamefont{P.~A.} \bibnamefont{Lee}} \bibnamefont{and}
  \bibinfo{author}{\bibfnamefont{T.~V.} \bibnamefont{Ramakrishnan}},
  \bibinfo{journal}{Rev. Mod. Phys.} \textbf{\bibinfo{volume}{57}},
  \bibinfo{pages}{287} (\bibinfo{year}{1985}).

\bibitem{Paalanen:1991_a}
\bibinfo{author}{\bibfnamefont{M.~A.} \bibnamefont{Paalanen}},
  \bibinfo{author}{\bibfnamefont{R.~N.} \bibnamefont{Bhatt}} \bibnamefont{and}
  \bibinfo{author}{\bibfnamefont{S.}~\bibnamefont{Sachdev}},
  \bibinfo{journal}{Physica B} \textbf{\bibinfo{volume}{169}},
  \bibinfo{pages}{223} (\bibinfo{year}{1991}).

\bibitem{Andrearczyk:2005_a}
\bibinfo{author}{\bibfnamefont{T.}~\bibnamefont{Andrearczyk}},
  \bibinfo{author}{\bibfnamefont{J.}~\bibnamefont{Jaroszy\'nski}},
  \bibinfo{author}{\bibfnamefont{G.}~\bibnamefont{Grabecki}},
  \bibinfo{author}{\bibfnamefont{T.}~\bibnamefont{Dietl}},
  \bibinfo{author}{\bibfnamefont{T.}~\bibnamefont{Fukumura}} \bibnamefont{and}
  \bibinfo{author}{\bibfnamefont{M.}~\bibnamefont{Kawasaki}},
  \bibinfo{journal}{Phys. Rev.} \textbf{\bibinfo{volume}{B 72}},
  \bibinfo{pages}{121309} (\bibinfo{year}{2005}).

\bibitem{Wojtowicz:1986_a}
\bibinfo{author}{\bibfnamefont{T.}~\bibnamefont{Wojtowicz}},
  \bibinfo{author}{\bibfnamefont{T.}~\bibnamefont{Dietl}},
  \bibinfo{author}{\bibfnamefont{M.}~\bibnamefont{Sawicki}},
  \bibinfo{author}{\bibfnamefont{W.}~\bibnamefont{Plesiewicz}}
  \bibnamefont{and}
  \bibinfo{author}{\bibfnamefont{J.}~\bibnamefont{Jaroszy\'nski}},
  \bibinfo{journal}{Phys. Rev. Lett.} \textbf{\bibinfo{volume}{56}},
  \bibinfo{pages}{2419} (\bibinfo{year}{1986}).

\bibitem{Jaroszynski:1992_a}
\bibinfo{author}{\bibfnamefont{J.}~\bibnamefont{{Jaroszy\'nski}}}
  \bibnamefont{and} \bibinfo{author}{\bibfnamefont{T.}~\bibnamefont{Dietl}},
  \bibinfo{journal}{Physica B} \textbf{\bibinfo{volume}{177}}
  \bibinfo{pages}{469} (\bibinfo{year}{1992}).

\bibitem{Wojtowicz:1989_a}
\bibinfo{author}{\bibfnamefont{H.}~\bibnamefont{Ohno}},
  \bibinfo{author}{\bibfnamefont{H.}~\bibnamefont{Munekata}},
  \bibinfo{author}{\bibfnamefont{T.}~\bibnamefont{Penney}},
  \bibinfo{author}{\bibfnamefont{S.}~\bibnamefont{{von Moln\'{a}r}}}
  \bibnamefont{and} \bibinfo{author}{\bibfnamefont{L.~L.} \bibnamefont{Chang}},
  \bibinfo{journal}{Physica B} \textbf{\bibinfo{volume}{155}},
  \bibinfo{pages}{357} (\bibinfo{year}{1989}).

\bibitem{Glod:1994_a}
\bibinfo{author}{\bibfnamefont{P.}~\bibnamefont{G\l\'od}},
  \bibinfo{author}{\bibfnamefont{T.}~\bibnamefont{Dietl}},
  \bibinfo{author}{\bibfnamefont{T.}~\bibnamefont{Fromherz}},
  \bibinfo{author}{\bibfnamefont{G.}~\bibnamefont{Bauer}} \bibnamefont{and}
  \bibinfo{author}{\bibfnamefont{I.}~\bibnamefont{Miotkowski}},
  \bibinfo{journal}{Phys. Rev. B} \textbf{\bibinfo{volume}{49}},
  \bibinfo{pages}{7797} (\bibinfo{year}{1994}).

\bibitem{Matsukura:1998_a}
\bibinfo{author}{\bibfnamefont{F.}~\bibnamefont{Matsukura}},
  \bibinfo{author}{\bibfnamefont{H.}~\bibnamefont{Ohno}},
  \bibinfo{author}{\bibfnamefont{A.}~\bibnamefont{Shen}} \bibnamefont{and}
  \bibinfo{author}{\bibfnamefont{Y.}~\bibnamefont{Sugawara}},
  \bibinfo{journal}{Phys. Rev. B} \textbf{\bibinfo{volume}{57}},
  \bibinfo{pages}{R2037} (\bibinfo{year}{1998}).

\bibitem{Benoit:1992_a}
\bibinfo{author}{\bibfnamefont{C.}~\bibnamefont{{Benoit \`a la Guillaume}}},
  \bibinfo{author}{\bibfnamefont{D.}~\bibnamefont{Scalbert}} \bibnamefont{and}
  \bibinfo{author}{\bibfnamefont{T.}~\bibnamefont{Dietl}},
  \bibinfo{journal}{Phys. Rev. B} \textbf{\bibinfo{volume}{46}},
  \bibinfo{pages}{9853(R)} (\bibinfo{year}{1992}).

\bibitem{Dietl:2002_a}
\bibinfo{author}{\bibfnamefont{T.}~\bibnamefont{Dietl}},
  \bibinfo{author}{\bibfnamefont{F.}~\bibnamefont{Matsukura}}
  \bibnamefont{and} \bibinfo{author}{\bibfnamefont{H.}~\bibnamefont{Ohno}},
  \bibinfo{journal}{Phys. Rev.} \textbf{\bibinfo{volume}{B 66}},
  \bibinfo{pages}{033203} (\bibinfo{year}{2002}).

\bibitem{Dietl:2007_b}
\bibinfo{author}{\bibfnamefont{T.}~\bibnamefont{Dietl}}, \eprint{cond-mat/0703278}.

\bibitem{Pacuski:2007_a}
\bibinfo{author}{\bibfnamefont{W.}~\bibnamefont{Pacuski}},
  \bibinfo{author}{\bibfnamefont{P.}~\bibnamefont{Kossacki}},
  \bibinfo{author}{\bibfnamefont{D.}~\bibnamefont{Ferrand}},
  \bibinfo{author}{\bibfnamefont{A.}~\bibnamefont{Golnik}},
  \bibinfo{author}{\bibfnamefont{J.}~\bibnamefont{Cibert}},
  \bibinfo{author}{\bibfnamefont{M.}~\bibnamefont{Wegscheider}},
  \bibinfo{author}{\bibfnamefont{A.}~\bibnamefont{Navarro-Quezada}},
  \bibinfo{author}{\bibfnamefont{A.}~\bibnamefont{Bonanni}},
  \bibinfo{author}{\bibfnamefont{M.}~\bibnamefont{Kiecana}},
  \bibinfo{author}{\bibfnamefont{M.}~\bibnamefont{Sawicki}},
  \bibnamefont{et~al.}, \bibinfo{journal}{Phys. Rev. Lett.} p.
  \bibinfo{pages}{in press} (\bibinfo{year}{2008}), \eprint{arXiv:0708.3296}.

\bibitem{Wachter:1979_a}
\bibinfo{author}{\bibfnamefont{P.}~\bibnamefont{Wachter}}, in
  \emph{\bibinfo{booktitle}{Handbook on the Physics and Chemistry of Rare
  Earth}}, edited by
  \bibinfo{editor}{\bibfnamefont{J.}~\bibnamefont{K.~A.~Gschneidner}}
  \bibnamefont{and} \bibinfo{editor}{\bibfnamefont{L.}~\bibnamefont{Eyring}}
  (\bibinfo{publisher}{North-Holland, Amsterdam}, \bibinfo{year}{1979}),
  vol.~\bibinfo{volume}{2}, p. \bibinfo{pages}{507}.

\bibitem{Dietl:1994_a}
\bibinfo{author}{\bibfnamefont{T.}~\bibnamefont{Dietl}}, in
  \emph{\bibinfo{booktitle}{Handbook of Semiconductors}}, edited by
  \bibinfo{editor}{\bibfnamefont{S.}~\bibnamefont{Mahajan}}
  (\bibinfo{publisher}{North Holland, Amsterdam}, \bibinfo{year}{1994}),
  vol.~\bibinfo{volume}{3B}, p. \bibinfo{pages}{1251}.

\bibitem{Jaczynski:1978_a}
\bibinfo{author}{\bibfnamefont{M.}~\bibnamefont{Jaczy\'nski}},
  \bibinfo{author}{\bibfnamefont{J.}~\bibnamefont{Kossut}} \bibnamefont{and}
  \bibinfo{author}{\bibfnamefont{R.~R.} \bibnamefont{Ga{\l}{\c{a}}zka}},
  \bibinfo{journal}{phys. stat. sol. (b)} \textbf{\bibinfo{volume}{88}},
  \bibinfo{pages}{73} (\bibinfo{year}{1978}).

\bibitem{Shapira:1974_a}
\bibinfo{author}{\bibfnamefont{Y.}~\bibnamefont{Shapira}} \bibnamefont{and}
  \bibinfo{author}{\bibfnamefont{R.~L.} \bibnamefont{Kautz}},
  \bibinfo{journal}{Phys. Rev. B} \textbf{\bibinfo{volume}{10}},
  \bibinfo{pages}{4781} (\bibinfo{year}{1974}).

\bibitem{Lopez-Sancho:2003_b}
\bibinfo{author}{\bibfnamefont{M.}~\bibnamefont{{L\'{o}pez-Sancho}}}
  \bibnamefont{and} \bibinfo{author}{\bibfnamefont{L.}~\bibnamefont{Brey}},
  \bibinfo{journal}{Phys. Rev.} \textbf{\bibinfo{volume}{B 68}},
  \bibinfo{pages}{113201} (\bibinfo{year}{2003}).

\bibitem{Hankiewicz:2004_c}
\bibinfo{author}{\bibfnamefont{E.~M.} \bibnamefont{Hankiewicz}},
  \bibinfo{author}{\bibfnamefont{T.}~\bibnamefont{Jungwirth}},
  \bibinfo{author}{\bibfnamefont{T.}~\bibnamefont{Dietl}},
  \bibinfo{author}{\bibfnamefont{C.}~\bibnamefont{Timm}} \bibnamefont{and}
  \bibinfo{author}{\bibfnamefont{J.}~\bibnamefont{Sinova}},
  \bibinfo{journal}{Phys. Rev. B} \textbf{\bibinfo{volume}{70}},
  \bibinfo{pages}{245211} (\bibinfo{year}{2004}).

\bibitem{Chiba:2006_a}
\bibinfo{author}{\bibfnamefont{D.}~\bibnamefont{Chiba}},
  \bibinfo{author}{\bibfnamefont{M.}~\bibnamefont{Yamanouchi}},
  \bibinfo{author}{\bibfnamefont{F.}~\bibnamefont{Matsukura}},
  \bibinfo{author}{\bibfnamefont{T.}~\bibnamefont{Dietl}} \bibnamefont{and}
  \bibinfo{author}{\bibfnamefont{H.}~\bibnamefont{Ohno}},
  \bibinfo{journal}{Phys. Rev. Lett.} \textbf{\bibinfo{volume}{96}},
  \bibinfo{pages}{096602} (\bibinfo{year}{2006}).

\bibitem{Jaroszynski:1985_a}
\bibinfo{author}{\bibfnamefont{J.}~\bibnamefont{Jaroszy\'nski}}
  \bibnamefont{and} \bibinfo{author}{\bibfnamefont{T.}~\bibnamefont{Dietl}},
  \bibinfo{journal}{Solid State Commun.} \textbf{\bibinfo{volume}{55}},
  \bibinfo{pages}{492} (\bibinfo{year}{1985}).

\bibitem{Sheu:2007_a}
\bibinfo{author}{\bibfnamefont{B.~L.} \bibnamefont{Sheu}},
  \bibinfo{author}{\bibfnamefont{R.~C.} \bibnamefont{Myers}},
  \bibinfo{author}{\bibfnamefont{J.-M.} \bibnamefont{Tang}},
  \bibinfo{author}{\bibfnamefont{N.}~\bibnamefont{Samarth}},
  \bibinfo{author}{\bibfnamefont{D.~D.} \bibnamefont{Awschalom}},
  \bibinfo{author}{\bibfnamefont{P.}~\bibnamefont{Schiffer}} \bibnamefont{and}
  \bibinfo{author}{\bibfnamefont{M.~E.} \bibnamefont{Flatt\'e}},
  \textbf{\bibinfo{volume}{99}},
  \bibinfo{pages}{227205}(\bibinfo{year}{2008}).

\bibitem{Dagotto:2001_a}
\bibinfo{author}{\bibfnamefont{E.}~\bibnamefont{Dagotto}},
  \bibinfo{author}{\bibfnamefont{T.}~\bibnamefont{Hotta}} \bibnamefont{and}
  \bibinfo{author}{\bibfnamefont{A.}~\bibnamefont{Moreo}},
  \bibinfo{journal}{Phys. Rep.} \textbf{\bibinfo{volume}{344}},
  \bibinfo{pages}{1} (\bibinfo{year}{2001}).

\bibitem{Nagaev:2001_a}
\bibinfo{author}{\bibfnamefont{E.~L.} \bibnamefont{Nagaev}},
  \bibinfo{journal}{Phys. Rep.} \textbf{\bibinfo{volume}{346}},
  \bibinfo{pages}{387} (\bibinfo{year}{2001}).

\bibitem{Sawicki:1986_a}
\bibinfo{author}{\bibfnamefont{M.}~\bibnamefont{Sawicki}},
  \bibinfo{author}{\bibfnamefont{T.}~\bibnamefont{Dietl}},
  \bibinfo{author}{\bibfnamefont{J.}~\bibnamefont{Kossut}},
  \bibinfo{author}{\bibfnamefont{J.}~\bibnamefont{Igalson}},
  \bibinfo{author}{\bibfnamefont{T.}~\bibnamefont{Wojtowicz}}
  \bibnamefont{and}
  \bibinfo{author}{\bibfnamefont{W.}~\bibnamefont{Plesiewicz}},
  \bibinfo{journal}{Phys. Rev. Lett.} \textbf{\bibinfo{volume}{56}},
  \bibinfo{pages}{508} (\bibinfo{year}{1986}).

\bibitem{Dietl:1986_a}
\bibinfo{author}{\bibfnamefont{T.}~\bibnamefont{Dietl}},
  \bibinfo{author}{\bibfnamefont{L.}~\bibnamefont{\'Swierkowski}},
  \bibinfo{author}{\bibfnamefont{J.}~\bibnamefont{Jaroszyñski}},
  \bibinfo{author}{\bibfnamefont{M.}~\bibnamefont{Sawicki}} \bibnamefont{and}
  \bibinfo{author}{\bibfnamefont{T.}~\bibnamefont{Wojtowicz}},
  \bibinfo{journal}{Physica Scrripta T} \textbf{\bibinfo{volume}{14}},
  \bibinfo{pages}{29} (\bibinfo{year}{1986}).

\bibitem{Glod:1994_b}
\bibinfo{author}{\bibfnamefont{P.}~\bibnamefont{G\l\'od}},
  \bibinfo{author}{\bibfnamefont{T.}~\bibnamefont{Dietl}},
  \bibinfo{author}{\bibfnamefont{M.}~\bibnamefont{Sawicki}} \bibnamefont{and}
  \bibinfo{author}{\bibfnamefont{I.}~\bibnamefont{Miotkowski}},
  \bibinfo{journal}{Physica B} \textbf{\bibinfo{volume}{194-196}},
  \bibinfo{pages}{995} (\bibinfo{year}{1994}).

\bibitem{Terry:1992_a}
\bibinfo{author}{\bibfnamefont{I.}~\bibnamefont{Terry}},
  \bibinfo{author}{\bibfnamefont{T.}~\bibnamefont{Penney}},
  \bibinfo{author}{\bibfnamefont{S.}~\bibnamefont{von Moln\'ar}}
  \bibnamefont{and} \bibinfo{author}{\bibfnamefont{P.}~\bibnamefont{Becla}},
  \bibinfo{journal}{Phys. Rev. Lett.} \textbf{\bibinfo{volume}{69}},
  \bibinfo{pages}{1800} (\bibinfo{year}{1992}).

\bibitem{Jaroszynski:2007_a}
\bibinfo{author}{\bibfnamefont{J.}~\bibnamefont{Jaroszy\'{n}ski}},
  \bibinfo{author}{\bibfnamefont{T.}~\bibnamefont{Andrearczyk}},
  \bibinfo{author}{\bibfnamefont{G.}~\bibnamefont{Karczewski}},
  \bibinfo{author}{\bibfnamefont{J.}~\bibnamefont{Wr\'{o}bel}},
  \bibinfo{author}{\bibfnamefont{T.}~\bibnamefont{Wojtowicz}},
  \bibinfo{author}{\bibfnamefont{D.}~\bibnamefont{Popovi\'{c}}}
  \bibnamefont{and} \bibinfo{author}{\bibfnamefont{T.}~\bibnamefont{Dietl}},
  \bibinfo{journal}{Phys. Rev. B} \textbf{\bibinfo{volume}{76}},
  \bibinfo{eid}{045322} (\bibinfo{year}{2007}).

\bibitem{Smorchkova:1997_a}
\bibinfo{author}{\bibfnamefont{I.}~\bibnamefont{Smorchkova}},
  \bibinfo{author}{\bibfnamefont{N.}~\bibnamefont{Samarth}},
  \bibinfo{author}{\bibfnamefont{J.}~\bibnamefont{Kikkawa}} \bibnamefont{and}
  \bibinfo{author}{\bibfnamefont{D.}~\bibnamefont{Awschalom}},
  \bibinfo{journal}{Phys. Rev. Lett.} \textbf{\bibinfo{volume}{78}},
  \bibinfo{pages}{3571} (\bibinfo{year}{1997}).

\bibitem{Dietl:2007_d}
\bibinfo{author}{\bibfnamefont{T.}~\bibnamefont{Dietl}},
  \bibinfo{author}{\bibfnamefont{T.}~\bibnamefont{Andrearczyk}},
  \bibinfo{author}{\bibfnamefont{A.}~\bibnamefont{{Lipi\'nska}}},
  \bibinfo{author}{\bibfnamefont{M.}~\bibnamefont{Kiecana}},
  \bibinfo{author}{\bibfnamefont{M.}~\bibnamefont{Tay}} \bibnamefont{and}
  \bibinfo{author}{\bibfnamefont{Y.}~\bibnamefont{Wu}}, \bibinfo{journal}{Phys.
  Rev. B} \textbf{\bibinfo{volume}{76}}, \bibinfo{pages}{155312}
  (\bibinfo{year}{2007}).

\bibitem{Kechrakos:2005_a}
\bibinfo{author}{\bibfnamefont{D.}~\bibnamefont{Kechrakos}},
  \bibinfo{author}{\bibfnamefont{N.}~\bibnamefont{Papanikolaou}},
  \bibinfo{author}{\bibfnamefont{K.~N.} \bibnamefont{Trohidou}}
  \bibnamefont{and} \bibinfo{author}{\bibfnamefont{T.}~\bibnamefont{Dietl}},
  \bibinfo{journal}{Phys. Rev. Lett.} \textbf{\bibinfo{volume}{94}},
  \bibinfo{pages}{127201} (\bibinfo{year}{2005}).

\bibitem{Ferrand:2001_a}
\bibinfo{author}{\bibfnamefont{D.}~\bibnamefont{Ferrand}},
  \bibinfo{author}{\bibfnamefont{J.}~\bibnamefont{Cibert}},
  \bibinfo{author}{\bibfnamefont{A.}~\bibnamefont{Wasiela}},
  \bibinfo{author}{\bibfnamefont{C.}~\bibnamefont{Bourgognon}},
  \bibinfo{author}{\bibfnamefont{S.}~\bibnamefont{Tatarenko}},
  \bibinfo{author}{\bibfnamefont{G.}~\bibnamefont{Fishman}},
  \bibinfo{author}{\bibfnamefont{T.}~\bibnamefont{Andrearczyk}},
  \bibinfo{author}{\bibfnamefont{J.}~\bibnamefont{{Jaroszy\'{n}ski}}},
  \bibinfo{author}{\bibfnamefont{S.}~\bibnamefont{{Kole\'{s}nik}}},
  \bibinfo{author}{\bibfnamefont{T.}~\bibnamefont{Dietl}},
  \bibnamefont{et~al.}, \bibinfo{journal}{Phys. Rev. B}
  \textbf{\bibinfo{volume}{63}}, \bibinfo{pages}{085201}
  (\bibinfo{year}{2001}).

\bibitem{Fukuyama:1979_a} H. Fukuyama and K. Yosida, J. Phys. Soc. Japan \textbf{46}, 1522 (1979).

\bibitem{Shapira:1990_a}
\bibinfo{author}{\bibfnamefont{Y.}~\bibnamefont{Shapira}},
  \bibinfo{author}{\bibfnamefont{N.~F.} \bibnamefont{{Oliveira Jr.}}},
  \bibinfo{author}{\bibfnamefont{P.}~\bibnamefont{Becla}} \bibnamefont{and}
  \bibinfo{author}{\bibfnamefont{T.~Q.} \bibnamefont{Vu}},
  \bibinfo{journal}{Phys. Rev. B} \textbf{\bibinfo{volume}{41}},
  \bibinfo{pages}{5931} (\bibinfo{year}{1990}).


\bibitem{Matsukura:2004_b}
\bibinfo{author}{\bibfnamefont{F.}~\bibnamefont{Matsukura}},
  \bibinfo{author}{\bibfnamefont{M.}~\bibnamefont{Sawicki}},
  \bibinfo{author}{\bibfnamefont{T.}~\bibnamefont{Dietl}},
  \bibinfo{author}{\bibfnamefont{D.}~\bibnamefont{Chiba}} \bibnamefont{and}
  \bibinfo{author}{\bibfnamefont{H.}~\bibnamefont{Ohno}},
  \bibinfo{journal}{Physica E} \textbf{\bibinfo{volume}{21}},
  \bibinfo{pages}{1032} (\bibinfo{year}{2004}).

\bibitem{Dugaev:2001_a}  V.~K. Dugaev, A. Crépieux and P. Bruno, Phys. Rev. B \textbf{64}, 104411 (2001).

\bibitem{Sawicki:1988_a} M.~Sawicki and T. Dietl, in {\it Proceedings of the 19th International Conference on the Physics of Semiconductors}, edited by W. Zawadzki (Institute of Physics, Polish Academy of Sciences, Warsaw, 1988) p.~1217.

\bibitem{Cumings:2006_a} J.~Cumings, L. S. Moore, H. T. Chou, K. C. Ku, G. Xiang, S. A. Crooker, N. Samarth and D. Goldhaber-Gordon, Phys. Rev. Lett. \textbf{96}, 196404 (2006).


\bibitem{Potashnik:2001_a}
\bibinfo{author}{\bibfnamefont{S.~J.} \bibnamefont{Potashnik}},
  \bibinfo{author}{\bibfnamefont{K.~C.} \bibnamefont{Ku}},
  \bibinfo{author}{\bibfnamefont{S.~H.} \bibnamefont{Chun}},
  \bibinfo{author}{\bibfnamefont{J.~J.} \bibnamefont{Berry}},
  \bibinfo{author}{\bibfnamefont{N.}~\bibnamefont{Samarth}} \bibnamefont{and}
  \bibinfo{author}{\bibfnamefont{P.}~\bibnamefont{Schiffer}},
  \bibinfo{journal}{Appl. Phys. Lett.} \textbf{\bibinfo{volume}{79}},
  \bibinfo{pages}{1495} (\bibinfo{year}{2001}).

\bibitem{Campion:2003_b}
\bibinfo{author}{\bibfnamefont{R.~P.} \bibnamefont{Campion}},
  \bibinfo{author}{\bibfnamefont{K.~W.} \bibnamefont{Edmonds}},
  \bibinfo{author}{\bibfnamefont{L.~X.} \bibnamefont{Zhao}},
  \bibinfo{author}{\bibfnamefont{K.~Y.} \bibnamefont{Wang}},
  \bibinfo{author}{\bibfnamefont{C.~T.} \bibnamefont{Foxon}},
  \bibinfo{author}{\bibfnamefont{B.~L.} \bibnamefont{Gallagher}}
  \bibnamefont{and} \bibinfo{author}{\bibfnamefont{C.~R.}
  \bibnamefont{Staddon}}, \bibinfo{journal}{J. Cryst. Growth}
  \textbf{\bibinfo{volume}{251}}, \bibinfo{pages}{311} (\bibinfo{year}{2003}).

\bibitem{Dietl:2000_a}
\bibinfo{author}{\bibfnamefont{T.}~\bibnamefont{Dietl}},
  \bibinfo{author}{\bibfnamefont{H.}~\bibnamefont{Ohno}},
  \bibinfo{author}{\bibfnamefont{F.}~\bibnamefont{Matsukura}},
  \bibinfo{author}{\bibfnamefont{J.}~\bibnamefont{Cibert}} \bibnamefont{and}
  \bibinfo{author}{\bibfnamefont{D.}~\bibnamefont{Ferrand}}
  \bibinfo{journal}{Science} \textbf{\bibinfo{volume}{287}},
  \bibinfo{pages}{1019} (\bibinfo{year}{2000}).

\bibitem{Dietl:2001_b}
\bibinfo{author}{\bibfnamefont{T.}~\bibnamefont{Dietl}},
  \bibinfo{author}{\bibfnamefont{H.}~\bibnamefont{Ohno}}, \bibnamefont{and}
  \bibinfo{author}{\bibfnamefont{F.}~\bibnamefont{Matsukura}},
  \bibinfo{journal}{Phys. Rev. B} \textbf{\bibinfo{volume}{63}},
  \bibinfo{pages}{195205} (\bibinfo{year}{2001}).

\bibitem{Kepa:2003_a}
\bibinfo{author}{\bibfnamefont{H.}~\bibnamefont{{K\c{e}pa}}},
  \bibinfo{author}{\bibfnamefont{L.~V.} \bibnamefont{Khoi}},
  \bibinfo{author}{\bibfnamefont{C.~M.} \bibnamefont{Brown}},
  \bibinfo{author}{\bibfnamefont{M.}~\bibnamefont{Sawicki}},
  \bibinfo{author}{\bibfnamefont{J.~K.} \bibnamefont{Furdyna}}
  \bibinfo{author}{\bibfnamefont{T.~M.} \bibnamefont{{Giebu{\l}towicz}}},
  \bibnamefont{and} \bibinfo{author}{\bibfnamefont{T.}~\bibnamefont{Dietl}},
  \bibinfo{journal}{Phys. Rev. Lett.} \textbf{\bibinfo{volume}{91}},
  \bibinfo{pages}{087205} (\bibinfo{year}{2003}).

\bibitem{Mayr:2002_a}
\bibinfo{author}{\bibfnamefont{M.}~\bibnamefont{Mayr}},
  \bibinfo{author}{\bibfnamefont{G.}~\bibnamefont{Alvarez}} \bibnamefont{and}
  \bibinfo{author}{\bibfnamefont{E.}~\bibnamefont{Dagotto}},
  \bibinfo{journal}{Phys. Rev.} \textbf{\bibinfo{volume}{B 65}},
  \bibinfo{pages}{241202} (\bibinfo{year}{2002}).

\bibitem{Oiwa:1997_a}
\bibinfo{author}{\bibfnamefont{A.}~\bibnamefont{Oiwa}},
  \bibinfo{author}{\bibfnamefont{S.}~\bibnamefont{Katsumoto}},
  \bibinfo{author}{\bibfnamefont{A.}~\bibnamefont{Endo}},
  \bibinfo{author}{\bibfnamefont{M.}~\bibnamefont{Hirasawa}},
  \bibinfo{author}{\bibfnamefont{Y.}~\bibnamefont{Iye}},
  \bibinfo{author}{\bibfnamefont{H.}~\bibnamefont{Ohno}},
  \bibinfo{author}{\bibfnamefont{F.}~\bibnamefont{Matsukura}},
  \bibinfo{author}{\bibfnamefont{A.}~\bibnamefont{Shen}} \bibnamefont{and}
  \bibinfo{author}{\bibfnamefont{Y.}~\bibnamefont{Sugawara}},
  \bibinfo{journal}{Solid State Commun.} \textbf{\bibinfo{volume}{103}},
  \bibinfo{pages}{209} (\bibinfo{year}{1997}).

\bibitem{Jungwirth:2006_b}
\bibinfo{author}{\bibfnamefont{T.}~\bibnamefont{Jungwirth}},
  \bibinfo{author}{\bibfnamefont{K.~Y.} \bibnamefont{Wang}},
  \bibinfo{author}{\bibfnamefont{J.}~\bibnamefont{{Ma\v{s}ek}}},
  \bibinfo{author}{\bibfnamefont{K.~W.} \bibnamefont{Edmonds}},
  \bibinfo{author}{\bibfnamefont{J.}~\bibnamefont{{K\"{o}nig}}},
  \bibinfo{author}{\bibfnamefont{J.}~\bibnamefont{Sinova}},
  \bibinfo{author}{\bibfnamefont{M.}~\bibnamefont{Polini}},
  \bibinfo{author}{\bibfnamefont{N.}~\bibnamefont{Goncharuk}},
  \bibinfo{author}{\bibfnamefont{A.~H.} \bibnamefont{MacDonald}},
  \bibinfo{author}{\bibfnamefont{M.}~\bibnamefont{Sawicki}},
  \bibnamefont{et~al.}, \bibinfo{journal}{Phys. Rev. B}
  \textbf{\bibinfo{volume}{73}}, \bibinfo{pages}{165205}
  (\bibinfo{year}{2006}).

\bibitem{Timm:2005_a}
\bibinfo{author}{\bibfnamefont{C.}~\bibnamefont{Timm}},
  \bibinfo{author}{\bibfnamefont{M.~E.} \bibnamefont{Raikh}} \bibnamefont{and}
  \bibinfo{author}{\bibfnamefont{F.}~\bibnamefont{{von Oppen}}},
  \bibinfo{journal}{Phys. Rev. Lett.} \textbf{\bibinfo{volume}{94}},
  \bibinfo{pages}{036602} (\bibinfo{year}{2005}).

\bibitem{Omiya:2000_a}
\bibinfo{author}{\bibfnamefont{T.}~\bibnamefont{Omiya}},
  \bibinfo{author}{\bibfnamefont{F.}~\bibnamefont{Matsukura}},
  \bibinfo{author}{\bibfnamefont{T.}~\bibnamefont{Dietl}},
  \bibinfo{author}{\bibfnamefont{Y.}~\bibnamefont{Ohno}},
  \bibinfo{author}{\bibfnamefont{T.}~\bibnamefont{Sakon}},
  \bibinfo{author}{\bibfnamefont{M.}~\bibnamefont{Motokawa}} \bibnamefont{and}
  \bibinfo{author}{\bibfnamefont{H.}~\bibnamefont{Ohno}},
  \bibinfo{journal}{Physica} \textbf{\bibinfo{volume}{E 7}},
  \bibinfo{pages}{976} (\bibinfo{year}{2000}).

\bibitem{Yuldashev:2003_a}
\bibinfo{author}{\bibfnamefont{S.~U.} \bibnamefont{Yuldashev}},
  \bibinfo{author}{\bibfnamefont{H.}~\bibnamefont{Im}},
  \bibinfo{author}{\bibfnamefont{V.~S.} \bibnamefont{Yalishev}},
  \bibinfo{author}{\bibfnamefont{C.~S.} \bibnamefont{Park}},
  \bibinfo{author}{\bibfnamefont{T.~W.} \bibnamefont{Kang}},
  \bibinfo{author}{\bibfnamefont{S.}~\bibnamefont{Lee}},
  \bibinfo{author}{\bibfnamefont{Y.}~\bibnamefont{Sasaki}},
  \bibinfo{author}{\bibfnamefont{X.}~\bibnamefont{Liu}} \bibnamefont{and}
  \bibinfo{author}{\bibfnamefont{J.~K.} \bibnamefont{Furdyna}},
  \bibinfo{journal}{Appl. Phys. Lett.} \textbf{\bibinfo{volume}{82}},
  \bibinfo{pages}{1206} (\bibinfo{year}{2003}).

\bibitem{Goennenwein:2005_a}
\bibinfo{author}{\bibfnamefont{S.~T.~B.} \bibnamefont{Goennenwein}},
  \bibinfo{author}{\bibfnamefont{S.}~\bibnamefont{Russo}},
  \bibinfo{author}{\bibfnamefont{A.~F.} \bibnamefont{Morpurgo}},
  \bibinfo{author}{\bibfnamefont{T.~M.} \bibnamefont{Klapwijk}},
  \bibinfo{author}{\bibfnamefont{W.}~\bibnamefont{van Roy}} \bibnamefont{and}
  \bibinfo{author}{\bibfnamefont{J.}~\bibnamefont{de~Boeck}},
  \bibinfo{journal}{Phys. Rev. B} \textbf{\bibinfo{volume}{B}},
  \bibinfo{pages}{193306} (\bibinfo{year}{2005}).

\bibitem{Honolka:2007_a}
\bibinfo{author}{\bibfnamefont{J.}~\bibnamefont{Honolka}},
  \bibinfo{author}{\bibfnamefont{S.}~\bibnamefont{Masmanidis}},
  \bibinfo{author}{\bibfnamefont{H.~X.} \bibnamefont{Tang}},
  \bibinfo{author}{\bibfnamefont{D.~D.} \bibnamefont{Awschalom}}
  \bibnamefont{and} \bibinfo{author}{\bibfnamefont{M.~L.}
  \bibnamefont{Roukes}}, \bibinfo{journal}{Phys. Rev. B}
  \textbf{\bibinfo{volume}{75}}, \bibinfo{eid}{245310} (\bibinfo{year}{2007}).


\bibitem{Neumaier:2007_a} D. Neumaier, M. Schlapps, U. Wurstbauer, J. Sadowski, M. Reinwald, W. Wegscheider and D. Weiss, arXiv:0711.3378v2.


\bibitem{Dugaev:2001_b}
\bibinfo{author}{\bibfnamefont{V.~K.} \bibnamefont{Dugaev}},
  \bibinfo{author}{\bibfnamefont{P.}~\bibnamefont{Bruno}} \bibnamefont{and}
  \bibinfo{author}{\bibfnamefont{J.}~\bibnamefont{Barna\'s}},
  \bibinfo{journal}{Phys. Rev. B} \textbf{\bibinfo{volume}{64}},
  \bibinfo{pages}{144423} (\bibinfo{year}{2001}).

\bibitem{Katsumoto:1998_a}
\bibinfo{author}{\bibfnamefont{S.}~\bibnamefont{Katsumoto}},
  \bibinfo{author}{\bibfnamefont{A.}~\bibnamefont{Oiwa}},
  \bibinfo{author}{\bibfnamefont{Y.}~\bibnamefont{Iye}},
  \bibinfo{author}{\bibfnamefont{H.}~\bibnamefont{Ohno}},
  \bibinfo{author}{\bibfnamefont{F.}~\bibnamefont{Matsukura}},
  \bibinfo{author}{\bibfnamefont{A.}~\bibnamefont{Shen}} \bibnamefont{and}
  \bibinfo{author}{\bibfnamefont{Y.}~\bibnamefont{Sugawara}},
  \bibinfo{journal}{phys. stat. sol. (b)} \textbf{\bibinfo{volume}{205}},
  \bibinfo{pages}{115} (\bibinfo{year}{1998}).

\bibitem{Tang:2003_a} H.~X. Tang, R.~K. Kawakami, D.~D. Awschalom and M.~L. Roukes, Phys. Rev. Lett. \textbf{90}, 107201 (2003).

\bibitem{Sankowski:2007_a}
\bibinfo{author}{\bibfnamefont{P.}~\bibnamefont{Sankowski}},
  \bibinfo{author}{\bibfnamefont{P.}~\bibnamefont{Kacman}},
  \bibinfo{author}{\bibfnamefont{J.~A.} \bibnamefont{Majewski}}
  \bibnamefont{and} \bibinfo{author}{\bibfnamefont{T.}~\bibnamefont{Dietl}},
  \bibinfo{journal}{Phys. Rev. B} \textbf{\bibinfo{volume}{75}},
  \bibinfo{eid}{045306} (\bibinfo{year}{2007}).

\bibitem{Pappert:2006_a}
\bibinfo{author}{\bibfnamefont{K.}~\bibnamefont{Pappert}},
  \bibinfo{author}{\bibfnamefont{M.~J.} \bibnamefont{Schmidt}},
  \bibinfo{author}{\bibfnamefont{S.}~\bibnamefont{Humpfner}},
  \bibinfo{author}{\bibfnamefont{C.}~\bibnamefont{Ruster}},
  \bibinfo{author}{\bibfnamefont{G.~M.} \bibnamefont{Schott}},
  \bibinfo{author}{\bibfnamefont{K.}~\bibnamefont{Brunner}},
  \bibinfo{author}{\bibfnamefont{C.}~\bibnamefont{Gould}},
  \bibinfo{author}{\bibfnamefont{G.}~\bibnamefont{Schmidt}} \bibnamefont{and}
  \bibinfo{author}{\bibfnamefont{L.~W.} \bibnamefont{Molenkamp}},
  \bibinfo{journal}{Phys. Rev. Lett.} \textbf{\bibinfo{volume}{97}},
  \bibinfo{pages}{186402} (\bibinfo{year}{2006}).

\bibitem{Csontos:2005_a} M. Csontos, T. Wojtowicz, X. Liu, M. Dobrowolska, B. Jank\'o, J. K. Furdyna and G. Mih\'aly, Phys. Rev. Lett. \textbf{95}, 227203 (2005).


\bibitem{Dietl:2007_a}
\bibinfo{author}{\bibfnamefont{T.}~\bibnamefont{Dietl}}, \bibinfo{journal}{J.
  Phys.: Condens. Matter} \textbf{\bibinfo{volume}{19}},
  \bibinfo{pages}{165204} (\bibinfo{year}{2007}).

\bibitem{Dietl:2007_e}
\bibinfo{author}{\bibfnamefont{T.}~\bibnamefont{Dietl}}, \bibinfo{journal}{J.
  Appl. Phys.}  (\bibinfo{year}{2008}), \eprint{arXiv:0711.0343}.

\end{thebibliography}
\end{document}